\documentclass[11pt]{article}
\usepackage{graphicx}
\usepackage{dcolumn}
\usepackage{array, longtable}
\usepackage{epsfig}
\usepackage{amsmath}
\usepackage{colordvi}
\usepackage{hhline}
\usepackage[numbers,sort&compress]{natbib}
\graphicspath{{figures/}}

\newcommand{\BaBarYear}{2007}

\newcommand{\BABARConfNumber} {07/010}
\newcommand{\SLACPubNumber}{12729}
\newcommand{\LANLNumber} {0708.2097}

\input {babarsym}
\newcommand{\bei}{\begin{itemize}}
\newcommand{\eei}{\end{itemize}}
\newcommand{\beq}{\begin{equation}}
\newcommand{\eeq}{\end{equation}}
\newcommand{\beqn}{\begin{eqnarray}}
\newcommand{\eeqn}{\end{eqnarray}}
\newcommand{\beqns}{\begin{eqnarray*}}
\newcommand{\eeqns}{\end{eqnarray*}}

\newcommand{\intl}{\int\limits}

\newcommand{\pvec}{{\bf p}}
\def\ie{{\em i.e.}}

\def\eg{{\em e.g.}}

\def\I{{\rm Im}}
\def\KS{\ensuremath{K^0_S}}
\def\de{\Delta E}
\def\dt{\deltat}

\def\mprime{m^\prime}
\def\thetaprime{\theta^\prime}
\def\dmd{\Delta m_d}
\def\dmt{\Delta t}
\def\cat{c}
\def\TM{{\rm TM}}
\def\SCF{{\rm SCF}}

\def\fscfave{\kern 0.18em\overline{\kern -0.18em f}_{\rm SCF}}
\def\fscf{f_{\rm SCF}}
\def\fscfi{f_{{\rm SCF},i}}
\def\Qtag{q_{\rm tag}}
\def\Qtagi{q_{{\rm tag},i}}
\def\Atagqq{A_{q\bar q,\,\rm tag}}
\def\Atag{A_{{B^+,\,\rm tag}}}
\def\Atagj{A_{{B^+,\,\rm tag},j}}

\def\Abar{\kern 0.18em\overline{\kern -0.18em A}{}}
\def\Amptp{{\cal A}}

\def\Amptpbar{\kern 0.18em\overline{\kern -0.18em {\cal A}}}

\def\absAmptp{|\Amptp|}
\def\absAmptpbar{|\Amptpbar|}
\def\AmpAll{|{\cal A}^\pm(\dmt)|^2}

\def\detJ{|\det J|}
\def\detJi{|\det J_i|}
\def\Rscf{R_{\rm SCF}}
\def\rar{\rightarrow}

\newcommand\ph{\phantom}

\newcommand{\Kspipi}             {\mbox{$\KS\pip\pim$}}
\newcommand{\BztoKspipi}         {\mbox{$\Bz \to \Kspipi$}}
\newcommand{\BzbtoKspipi}        {\mbox{$\Bzb \to \Kspipi$}}

\def\AmpAllSigp{|{\cal A}_{\rm sig}^+(\dmt)|^2}
\def\AmpAllSigm{|{\cal A}_{\rm sig}^-(\dmt)|^2}
\def\AmpAllSigpm{|{\cal A}_{\rm sig}^\pm(\dmt)|^2}

\def\pipi   {\ensuremath{\pip\pim}\xspace}

\newcommand{\KstarI}             {\mbox{$\Kstar(892)$}}

\newcommand{\KstarpI}            {\mbox{$\Kstarp(892)$}}
\newcommand{\KstarmI}            {\mbox{$\Kstarm(892)$}}
\newcommand{\KstarpmI}           {\mbox{$\Kstarpm(892)$}}

\newcommand{\KstarpII}           {\mbox{$\Kstarp(1430)$}}

\newcommand{\KstarpmIV}          {\mbox{$\Kstarpm(1680)$}}

\newcommand{\rhoz}               {\mbox{$\rho^0$}}

\newcommand{\rhoI}               {\mbox{$\rhoz(770)$}}

\newcommand{\fI}                 {\mbox{$f_0(980)$}}

\newcommand{\fIKs}               {\mbox{$\fI \KS$}}

\newcommand{\rhoII}              {\mbox{$\rhoz(1450)$}}

\newcommand{\fII}                {\mbox{$f_2(1270)$}}

\newcommand{\fX}                {\mbox{$f_X(1300)$}}

\newcommand{\D}                  {\mbox{$D$}}

\newcommand{\half}{\mbox{$\frac{1}{2}$}}

\def\eg                 {{\it e.g.~}}

\def\ie                 {{\it i.e.~}}

\def\mBz{m_{\Bz}}
\def\spz{s_{+}}
\def\smz{s_{-}}
\def\spm{s_{0}}

\def\mpm{m_{0}}

\def\mpmMax{\mpm^{\rm max}}
\def\mpmMin{\mpm^{\rm min}}

\long\def\inst#1{\par\nobreak\kern 4pt\nobreak
    {\it #1}\par\vskip 10pt plus 3pt minus 3pt}

\begin{document}
{\pagestyle{empty}

\begin{flushright}
\babar-CONF-\BaBarYear/\BABARConfNumber \\
SLAC-PUB-\SLACPubNumber \\
hep-ex/\LANLNumber \\
August 2007 \\
\end{flushright}
\par\vskip 5cm

\begin{center}
{\Large \bf
\boldmath
Time-dependent Dalitz Plot Analysis of $B^0 \to \Kspipi$ 
} 
\end{center}
\bigskip

\begin{center}
\large The \babar\ Collaboration\\
\mbox{ }\\
\today
\end{center}
\bigskip \bigskip

\begin{center}
\large \bf Abstract
\end{center}
We perform a time-dependent Dalitz plot analysis of $B^0 \to \Kspipi$
in order to extract the  \CP violation parameters of
$\fI \KS$ and $\rhoI \KS$ and direct \CP-asymmetries of $\KstarpI \pi^-$. The results are obtained from a data sample 
of $(383\pm3)\times10^{6}$ \BB\ decays, collected 
with the \babar\ detector at the \pep2\ asymmetric--energy \B\ factory at SLAC.
The measured values of $2\beta_{\rm eff}$ in $\Bz$ decays to $\fI\KS$ and $\rhoI\KS$ are 
$(89^{+22}_{-20} \pm 5 \pm 8)^\circ$ and $(37 ^{+19}_{-17} \pm 5 \pm 6)^\circ$, respectively, 
where the first quoted uncertainty is statistical, the second is systematic and the third  is Dalitz plot signal 
model uncertainty.  We measure the significance of $2\beta_{\rm eff}(\fI\KS) \neq 0$ to be $4.3\,\sigma$.
In decays to $\KstarI \pi$ we find $A_{\CP} = -0.18 \pm 0.10 \pm 0.03 \pm 0.03$. 
The measured phase difference between the decay amplitudes of $\Bz \to \KstarpI \pim$ and $\Bzb \to \KstarmI \pip$ 
is $(-164 \pm 24 \pm 12 \pm 15)^\circ$.  All results are preliminary. 

\vfill
\begin{center}
Contributed to the 
XXIII$^{\rm rd}$ International Symposium on Lepton and Photon Interactions at High~Energies, 8/13 -- 8/18/2007, Daegu, Korea
\end{center}

\vspace{1.0cm}
\begin{center}
{\em Stanford Linear Accelerator Center, Stanford University, 
Stanford, CA 94309} \\ \vspace{0.1cm}\hrule\vspace{0.1cm}
Work supported in part by Department of Energy contract DE-AC03-76SF00515.
\end{center}

\newpage
}

\begin{center}
\small

The \babar\ Collaboration,
\bigskip

%
{B.~Aubert,}
{M.~Bona,}
{D.~Boutigny,}
{Y.~Karyotakis,}
{J.~P.~Lees,}
{V.~Poireau,}
{X.~Prudent,}
{V.~Tisserand,}
{A.~Zghiche}
\inst{Laboratoire de Physique des Particules, IN2P3/CNRS et Universit\'e de Savoie, F-74941 Annecy-Le-Vieux, France }
{J.~Garra~Tico,}
{E.~Grauges}
\inst{Universitat de Barcelona, Facultat de Fisica, Departament ECM, E-08028 Barcelona, Spain }
% ...

\end{center}\newpage

\setcounter{footnote}{0}


\section{INTRODUCTION}
\label{sec:introduction}

The Cabibbo-Kobayashi-Maskawa (CKM)
mechanism~\cite{Cabibbo:1963yz,Kobayashi:1973fv} for quark mixing
describes all transitions between quarks in terms of only four
parameters: three real rotation angles and one irreducible phase.
Consequently, the flavor sector of the Standard Model (SM)
is highly predictive.  One particularly interesting prediction is that
mixing-induced \CP asymmetries in decays governed by $b \to
q\bar{q}s$ ($q = u,d,s$) transitions are, to a good
approximation, the same as those found in $b \to c\bar{c}s$
transitions.  Since flavor changing neutral currents are forbidden at
tree-level in the Standard Model, the $b \to s$ transition proceeds
via loop diagrams (penguins), which are affected by new particles in
many extensions of the SM.

Recently, various different $b \to s$ dominated charmless hadronic $B$
decays have been studied in order to probe this prediction.  The
values of the mixing-induced \CP asymmetry measured for each
(quasi-)two-body mode can be compared to that measured in $b \to
c\bar{c}s$ transitions (typically using $B^0 \to J/\psi \KS$).  A
recent compilation~\cite{hfag} of these results 
shows that they tend to have
central values below that for $b \to c\bar{c}s$. However, there is currently no
convincing evidence for new physics effects in these transitions.
The most recent theoretical
evaluations~\cite{Grossman:2003qp,Gronau:2003kx,Gronau:2004hp,Cheng:2005bg,Gronau:2005gz,Beneke:2005pu,Engelhard:2005hu,Cheng:2005ug,Williamson:2006hb}
suggest that SM corrections to the $b \to q\bar{q}s$ mixing-induced
\CP violation parameters should be small (in particular for the modes
$\phi K^0$, $\eta^\prime K^0$ and $\KS\KS\KS$), and tend to
{\it increase} the values (\ie the opposite trend to that seen
in the data). Clearly, more precise experimental results are required.

The compilation given in
\cite{hfag}
includes several
three-body modes, which may be used either by virtue of being \CP
eigenstates ($\KS\KS\KS$, $\KS\pi^0\pi^0$)~\cite{Gershon:2004tk} or
since their \CP content can be determined experimentally
($K^+K^-K^0$)~\cite{Garmash:2003er,kkks:2007sd}. 
It also includes quasi-two-body (Q2B) modes, such as $f_0(980)\KS$ and
$\rhoI\KS$, which are reconstructed via three-body final states
($\KS\pi^+\pi^-$ for these modes).  For these channels, the precision
of the Q2B approach is limited as other structures in the
phase space 
may cause interference with the resonances
considered as signal.  Therefore, more precise results can be obtained
using a full, time-dependent Dalitz plot (DP) fit of $B^0 \to
\KS\pi^+\pi^-$.   Furthermore the
interference terms allow the cosine of the effective weak phase
difference in mixing and decay to be determined, helping to
resolve ambiguities which arise from the Q2B analysis.  This approach
has recently been successfully used in a time-dependent Dalitz plot
analysis of $\Bz \to K^+K^-K^0$~\cite{kkks:2007sd}.

The discussion above assumes that the $b \to s$ penguin amplitude
dominates the decay.  However, for each mode contributing to the
$\KS\pi^+\pi^-$ final state, there is also the possibility of a $b \to
u$ tree diagram.  These are doubly CKM suppressed compared to the $b
\to s$ penguin (the tree is ${\cal O}(\lambda^4)$ whereas the penguin
is  ${\cal O}(\lambda^2)$, where $\lambda$ is the usual Wolfenstein
parameter~\cite{Wolfenstein:1983yz,Buras:1994ec}).  However, hadronic
factors may result in a relative enhancement and hence significant
``tree pollution''.  The relative magnitudes of the tree and penguin
amplitudes, $\left| T/P \right|$, thus can be different for each Q2B
state, as can the strong phase difference.  However, the relative weak
phase between them is of course the same -- and in the Standard Model
is approximately equal to $\gamma$.  An amplitude analysis, in contrast
to a Q2B analysis, yields sufficient information to extract relative
phases and magnitudes.  DP analysis of  $B^0 \to
\KS\pi^+\pi^-$ (and similar modes) can therefore be used to determine
$\gamma$~\cite{Deshpande:2002be,Ciuchini:2006kv,Gronau:2006qn,Lipkin:1991st}.  A comparison of the
value obtained with that extracted from tree-level $B \to DK$ decays
provides a SM test. 

No results on time-dependent DP analysis of  $B^0 \to \Kspipi$ have
yet been published.
Belle have presented results of an
analysis using 388 million $\BB$ pairs~\cite{Garmash:2006fh},
which does not take into account either time-dependence or flavour-tag dependence.
The results of the Belle analysis are
consistent with other studies of the $B^0 \to \Kspipi$
decay~\cite{Garmash:2003er,kelly-paper}, as well as with measurements
obtained from other $K\pi\pi$ systems:
$K^+\pi^-\pi^0$~\cite{Chang:2004um,Aubert:2004bt} and
$K^+\pi^+\pi^-$~\cite{Aubert:2005ce,Garmash:2005rv}.
The latter results indicate evidence for direct \CP violation in the
$B^+ \to \rho^0 K^+$ channel.  If confirmed, this will be the first
observation of \CP violation in the decay of any charged
particle. Taken together with the observation of direct \CP violation
in $B^0 \to K^+\pi^-$ decays~\cite{Aubert:2004qm,Abe:2005fz}, these
results suggest that large \CP violation effects are possible in $B^0
\to K^{*+}\pi^-$ (although current measurements of the effect are
consistent with zero~\cite{kelly-paper}).

In this paper we present preliminary results from the first
time-dependent Dalitz plot analysis of the $\BztoKspipi$ decay.
In Section~\ref{sec:ana_overview} we describe the time-dependent Dalitz plot
formalism, and introduce the signal parameters that are extracted in our
fit to data.  In Section~\ref{sec:babar} we briefly describe the \babar\ detector and
the data set.  In Section~\ref{subsec:selection}, we explain the selection requirements used
to obtain our signal candidates and suppress backgrounds. In the same section we describe the
methods used to control experimental effects such as resolution in the
fit to data.  In Section~\ref{sec:fitResults} we present the results of the fit,
and extract parameters relevant to the contributing Q2B decays.  In
Section~\ref{sec:Systematics} we discuss systematic uncertainties in our results, and finally we
summarize our results in Section~\ref{sec:Summary}.

\section{ANALYSIS OVERVIEW}
\label{sec:ana_overview}

Using a maximum-likelihood fit, we measure relative phases and magnitudes for the
different resonant decay modes, taking advantage of the interference between them
in the Dalitz plot. Below we detail the formalism used in the present analysis.

\subsection{DECAY AMPLITUDES}
\label{sec:kinematics}

We consider the decay of a spin-zero $\Bz$ with four-momentum
$p_B$ into the three daughters $\pip$, $\pim$ and $\KS$
with $p_+$, $p_-$ and $p_0$, their corresponding four-momenta. Using
as independent (Mandelstam) variables the invariant squared masses
\beq
\label{eq:dalitzVariables}
       \spz \;=\; (p_+ + p_0)^2~, \hspace{1cm}
       \smz \;=\; (p_- + p_0)^2~, 
\eeq
the invariant squared mass of the positive and negative pion, 
$\spm \;=\; (p_+ + p_-)^2$, is obtained from energy and 
momentum conservation
\beq
\label{eq:magicSum}
        \spm \;=\; \mBz^2 + 2m_{\pi^+}^2 + m_{\KS}^2
                   - \spz - \smz~.
\eeq
The differential $\Bz$ decay width with respect to the 
variables defined in Equation~(\ref{eq:dalitzVariables}) (\ie the 
{\em Dalitz plot}) reads
\beq
\label{eq:partialWidth}
        d\Gamma(\BztoKspipi) \;=\; 
        \frac{1}{(2\pi)^3}\frac{|\Amptp|^2}{32 \mBz^3}\,d\spz d\smz~,
\eeq
where $\Amptp$ is the Lorentz-invariant amplitude
of the three-body decay. 
In the following, the amplitudes ${\cal A}$ and its \CP conjugate $\overline{\cal A}$ correspond to the transitions $\BztoKspipi$ and $\BzbtoKspipi$, respectively.
We describe the distribution of signal events 
in the DP using an isobar approximation,
which models the total amplitude as
resulting from a sum of amplitudes from the individual decay channels
\begin{eqnarray}
  \label{eq:isobar}
  {\cal A}(\spz,\smz) 
  & = & \sum_{j=1}^{N} c_j F_j(\spz,\smz) \\
  \overline{\cal A}(\spz,\smz) 
  & = & \sum_{j=1}^{N} \overline{c}_j \overline{F}_j(\spz,\smz)
\end{eqnarray}
where $F_j$ are DP dependent dynamical amplitudes described in the following, 
and $c_j$ are complex coefficients describing the relative
magnitude and phase of the different decay channels.
All the weak phase dependence is contained in $c_j$, and $F_j$
contains strong dynamics only, therefore 
\begin{eqnarray}
  F_j(\spz,\smz) & = & \overline{F}_j(\smz,\spz) ~.
\end{eqnarray}
The resonance dynamics are contained within the $F_j$ terms, which are represented
by the product of the invariant mass and angular distribution probabilities, \ie
\begin{equation}
\label{eq:ResDynEqn}
F_j^L(\spz,\smz) = R_j(m) \times X_L(|\vec{p}\,^{\star}|\,r) \times X_L(|\vec{q}\,|\,r) \times T_j(L,\vec{p},\vec{q}\,)~,
\end{equation}
where
\begin{itemize}
\item $m$ is the invariant mass of the decay products of the resonance,
\item $R_j(m)$ is the resonance mass term or ``lineshape'' (e.g.~Breit--Wigner),
\item $X_L$ are Blatt--Weisskopf barrier factors~\cite{blatt-weisskopf} with parameter $r$. These factors are taken to be unity in the present analysis, both in Equation~\eqref{eq:ResDynEqn} and in the lineshapes. The effect of this choice is accounted for as a systematic uncertainty.
\item $\vec{p}\,^{\star}$ is the momentum of the bachelor particle
      evaluated in the rest frame of the $B$,
\item $\vec{p}$ and $\vec{q}$ are the momenta of the bachelor particle and
      one of the resonance daughters respectively, both evaluated in the
      rest frame of the resonance,
\item $L$ is the orbital angular momentum between the resonance and the
      bachelor, and
\item $T_j(L,\vec{p},\vec{q})$ is the angular distribution, where:
\begin{eqnarray}
L=0 &:& T_j = 1~,\\
L=1 &:& T_j = -4\vec{p}\cdot\vec{q}~,\\
L=2 &:& T_j = \frac{8}{3} \left[3(\vec{p}\cdot\vec{q}\,)^2 - (|\vec{p}\,||\vec{q}\,|)^2\right]~.
\end{eqnarray}
\end{itemize}

The lineshape differs for each component included in the fit. The lineshapes used are Relativistic Breit--Wigner (RBW)~\cite{pdg2006}, Flatt\'e~\cite{Flatte}, Gounaris-Sakurai (GS)~\cite{Gounaris:1968mw} and  LASS~\cite{Aubert:2005ce,LASS,bugg}. A flat phase space term has been included in the signal model to account for nonresonant (NR) $\Bz \to \Kspipi$ decays. The components of the signal model are summarized in Table \ref{tab:model}. 

\begin{table}[htb]
\begin{center}
\begin{tabular}{cccc}
\hline\hline
Resonance      & Parameters                     & Form Factor & Ref. for         \\
               &                                &             & Parameters       \\ \hline
$f_0$          & $\text{mass}=965 \pm 10$       & Flatt\'e    & \cite{valFlatte} \\
               & $g_{\pi}=165 \pm 18$           &             &                  \\
               & $g_{K}=695 \pm 93$             &             &                  \\ \hline 
$\rho^0$       & $\text{mass}=775.5 \pm 0.4$    & GS          & \cite{pdg2006}       \\
               & $\text{width}=146.4 \pm 1.1$   &             &                  \\ \hline
$K^{*+}(892)$  & $\text{mass}=891.66 \pm 0.26$  & RBW         & \cite{pdg2006}       \\
$K^{*-}(892)$  & $\text{width}=50.8 \pm 0.9$    &             &                  \\ \hline
$K^{*+}(1430)$ & $\text{mass}=1415 \pm 3$       & LASS        & \cite{valLASS,Aubert:2005ce}   \\
$K^{*-}(1430)$ & $\text{width}=300 \pm 6$       &             &                  \\
               & $\text{cutoff}=2000$           &             &                  \\ 
               & $a=2.07\pm0.1\,({\rm GeV}^{-1})$       &     &                  \\
               & $r=3.32\pm0.34\,({\rm GeV}^{-1})$      &     &                  \\ \hline
$f_X(1300)$    & $\text{mass}=1449 \pm 13$      & RBW         & \cite{Garmash:2006fh}\\
               & $\text{width}=126\pm25$        &             &                  \\ \hline
$f_2(1270)$    & $\text{mass}= 1275.4\pm1.1$    & RBW         & \cite{pdg2006}   \\
               & $\text{width}=185.2^{+3.1}_{-2.5}$  &        &                  \\ \hline
$\chi_{c0}(1P)$& $\text{mass}=3414.75\pm0.35$   & RBW         & \cite{pdg2006}   \\
               & $\text{width}=10.4 \pm 0.7$    &             &                  \\ \hline
NR decays      &                                & flat phase space &             \\
\hline \hline
\end{tabular}
\end{center}
\caption{\em Parameters of the DP model used in the fit. Values are given in  ${\rm MeV}$, unless mentioned otherwise. 
\label{tab:model}}
\end{table}

\subsection{TIME DEPENDENCE}

With $\deltat \equiv t_{\rm sig} - t_{\rm tag}$ defined as the proper 
time interval between the decay of the fully reconstructed $\BztoKspipi$
($B^0_{{\rm sig}}$)
and that of the  other meson ($\Bz_{\rm tag}$) from the \FourS,  the time-dependent decay
rate $\AmpAllSigp$ ($\AmpAllSigm$) when the $\Bz_{\rm tag}$ is a $\Bz$ ($\Bzb$) 
is given by 
\beqn
\label{eq:dt}
    \AmpAllSigpm&
        =&
                \frac{e^{-|\dmt|/\tau_{B^0}}}{4\tau_{B^0}}
        \bigg[\absAmptp^2 + \absAmptpbar^2\nonumber\\
&&            \mp \left(\absAmptp^2 - \absAmptpbar^2\right)\cos(\dmd\dmt)\nonumber\\
&&            \pm\,2\I\left[\Amptpbar\Amptp^*\right]\sin(\dmd\dmt)   
        \bigg]~,
\eeqn
where $\tau_{B^0}$ is the  neutral $B$ meson lifetime and $\deltamd$ is the $\BzBzb$ 
mass difference. In the last formula and in the following, the DP dependence 
of amplitudes is implicit.
Here, we have assumed that 
there is no \CP violation in mixing, and have used a
convention whereby the phase from $\BzBzb$ mixing is absorbed into
the $\Bzb$ decay amplitude~\footnote{In other terms, we assume that $|q/p|=1$ and absorb $q/p$ into $\bar{c}_j$.} (\ie into the $\bar{c}_j$ terms).
Lifetime differences in the neutral $B$ meson system are assumed to be
negligible.

\subsection{THE SQUARE DALITZ PLOT}
\label{sec:SquareDP}

Both the signal events and the combinatorial $\epem\to q\bar q$ ($q=u,d,s,c$) 
continuum background events populate the kinematic boundaries of the 
DP due to the low final state masses compared with the $\Bz$ mass. 
We find the representation in Equation~(\ref{eq:partialWidth}) is inconvenient when one wants 
to use empirical reference shapes in a maximum-likelihood fit.
Large variations occurring in small areas of the DP are very difficult to describe in detail.
We therefore apply the transformation
\beq
\label{eq:SqDalitzTrans}
        d\spz \,d\smz \;\longrightarrow \detJ\, d\mprime\, d\thetaprime~,
\eeq
which defines the {\em Square Dalitz plot} (SDP). The new coordinates 
are
\beq
\label{eq:SqDalitzVars}
        \mprime \equiv \frac{1}{\pi}
                \arccos\left(2\frac{\mpm - \mpmMin}{\mpmMax - \mpmMin}
                        - 1
                      \right),~
        \thetaprime \equiv \frac{1}{\pi}\theta_{0}~,
\eeq
where $\mpm=\sqrt{s_0}$ is the $\pip\pim$ invariant mass,
$\mpmMax=\mBz - m_{\KS}$ and $\mpmMin=2m_{\pi^+}$ are the kinematic
limits of $\mpm$, $\theta_{0}$ is the
$\pi^+ \pi^-$ resonance
helicity angle, defined as the angle between the $\pim$ and the $\KS$ in the 
$\pi^+ \pi^-$ rest frame,
and $J$ is the Jacobian of the transformation.
Both variables  range between 0 and 1.
The determinant of the Jacobian is given by
\beq
\label{eq:detJ}
        \detJ \;=\;     4 \,|{\bf p}^*_+||{\bf p}^*_0| \,\mpm
                        \cdot   
                        \frac{\partial \mpm}{\partial \mprime}
                        \cdot   
                        \frac{\partial \cos\theta_{0}}{\partial \thetaprime}~,
\eeq
where 
$|{\bf p}^*_+|=\sqrt{E^{*\,2}_+ - m_{\pi^+}^2}$ and
$|{\bf p}^*_0|=\sqrt{E^{*\,2}_0 - m_{\KS}^2}$, and where the $\pi^+$ ($\KS$) energy 
$E^*_+$ ($E^*_0$), is defined in the $\pi^+\pi^-$ rest frame.
This same transformation has been used in previous $B$ decay DP 
analyses, \eg~Ref.~\cite{rhopipaper}.

\section{THE \babar\ DETECTOR AND DATASET}
\label{sec:babar}

The data used in this analysis were collected with the \babar\ 
detector at the \pep2\ asymmetric-energy $e^+e^-$ storage ring at 
SLAC between October 1999 and August 2006. The sample consists of about
$347\;\mathrm{fb}^{-1}$, corresponding to $(383\pm3)\times10^{6}$ 
$B\Bbar$ pairs collected at the \FourS resonance (``on-resonance''), 
and an integrated luminosity of $36.6$~\invfb collected about $40$~\mev 
below the~\FourS (``off-resonance'').

A detailed description of the \babar\ detector is presented in 
Ref.~\cite{babar}. The tracking system used for track and vertex 
reconstruction has two  components: a silicon vertex tracker 
(SVT) and a drift chamber (DCH), both operating within a 1.5~T 
magnetic field generated by a superconducting solenoidal magnet. 
Photons are identified in an electromagnetic calorimeter (EMC) 
surrounding a detector of internally reflected Cherenkov light 
(DIRC), which associates Cherenkov photons with tracks for particle 
identification (PID). Muon candidates are identified with the
use of the instrumented flux return (IFR) of the solenoid.

\section{EVENT SELECTION AND BACKGROUND SUPPRESSION}
\label{subsec:selection}

We reconstruct $\BztoKspipi$ candidates 
from pairs of 
oppositely-charged tracks and a $\KS\to\pip\pim$ candidate, which are required to form a good quality vertex.
In order to ensure that all events are within 
the DP boundaries, we constrain the  invariant mass of the final state to the $B$ mass.  
For the $\pi^+\pi^-$ pair from the $B$, 
we use information  from the tracking system, EMC, and DIRC to 
remove tracks consistent with electron, kaon and proton hypotheses.  
In addition, we require at least one track to be inconsistent with
the muon hypothesis based on information from the IFR.
The $\KS$ candidate is required to have a mass within $15\mevcc$ of 
the nominal $K^0$ mass \cite{pdg2006} and a decay vertex well separated 
from the $B^0$ decay vertex. In addition, the cosine 
of the angle between the $\KS$ flight direction and the vector between 
the $B$-daughter pions and the $\KS$ vertices must be greater than $0.999$.

A $B$-meson candidate is characterized kinematically by the energy-substituted 
mass $\mes\equiv\sqrt{(s/2+{\mathbf {p}}_i\cdot{\mathbf{p}}_B)^2/E_i^2-{\mathbf {p}}_B^2}$
and energy difference $\de \equiv E_B^*-\half\sqrt{s}$, 
where $(E_B,\pvec_B)$ and $(E_i,\pvec_i)$ are the four-vectors
of the $B$-candidate and the initial electron-positron system,
respectively. The asterisk denotes the \FourS\  frame,
and $s$ is the square of the invariant mass of the electron-positron system.  
We require $5.272 < \mes <5.286\gevcc$ and $|\de|<0.065\gev$.

Backgrounds arise primarily from random combinations in continuum events.
To enhance discrimination between signal and continuum, we 
use a neural network (NN)~\cite{NNo} to combine four discriminating variables: 
the angles with respect to the beam axis of the $B$ momentum and $B$ thrust 
axis in the \FourS\ frame; and the zeroth and second order monomials
$L_{0,2}$ of the energy flow about the $B$ thrust axis.  The monomials
are defined by $ L_n = \sum_i {\bf p}_i\times\left|\cos\theta_i\right|^n$,
where $\theta_i$ is the angle with respect to the $B$ thrust axis of
track or neutral cluster $i$, ${\bf p}_i$ is its momentum, and the sum
excludes the $B$ candidate.  
The NN is trained using off-resonance data as well as
simulated signal events that passed the selection criteria.
The final sample of signal candidates 
is selected with a requirement on the NN output that retains $90\%$ ($29\%$) 
of the signal (continuum).

The time difference $\deltat$ is obtained from the measured distance between 
the $z$ positions (along the beam direction) of the $\Bz_{\rm sig}$ and 
$\Bz_{\rm tag}$ decay vertices, and the boost $\beta\gamma=0.56$ of 
the \epem\ system: $\deltat = \Delta z/\beta\gamma c$.
$\Bz$ candidates with $|\deltat|>20$~ps are rejected, as well as candidates
for which the error on $\deltat$ is higher than $2.5$~ps.
To determine the flavor of the $\Bz_{\rm tag}$ 
we use the $B$ flavor tagging algorithm of Ref.~\cite{BabarS2b}.
This produces six mutually exclusive tagging categories. We also 
retain untagged events in a seventh category to improve the efficiency 
of the signal selection and because these events contribute to the 
measurement of direct \CP violation~\cite{Gardner:2003su}. Events with multiple \B 
candidates passing the full selection occur 
between $\sim1\%$ of the time for nonresonant signal and $\sim8\%$
of the time for $\Bz \to \fIKs$ signal.
If an event has more than one candidate, 
we select one using an arbitrary but reproducible procedure based
on the event timestamp. 

With the above selection criteria, we  obtain a signal efficiency determined 
from Monte Carlo (MC) simulation of $21-25\%$ depending on the composition of the 
DP.  

Of the selected signal events, $8\%$ of $B^0 \to \rho^0\KS$, 
$6\%$ of $B^0 \to \KstarI^+\pim$ and $4\% $ of $B^0 \to \fI\KS$ events are
misreconstructed.  Misreconstructed events occur when a track
from the tagging $B$ is assigned to the reconstructed signal candidate. 
This occurs most often for  low-momentum tracks  and hence the misreconstructed events 
are concentrated in the corners of the DP.  Since these are also the areas where the low-mass resonances 
overlap strongly with other resonances, it is important to model the misreconstruced events correctly.  The details of the model
for misreconstructed events over the DP is detailed in Section \ref{sec:deltaT}.

\subsection{BACKGROUND FROM OTHER {\em B} DECAYS}

\begin{table*}[t]
\begin{center}
\setlength{\tabcolsep}{0.0pc}
\begin{tabular*}{\textwidth}{@{\extracolsep{\fill}}lccc}
\hline
 Mode                                    & Varied & BR   & Number of events\\
\hline\\[-0.3cm]
 $B^0 \rar \D^-(\to\KS\pim)\pip$         & yes & --                              & $ 3377 \pm 60  $ \\
 $B^0 \rar \jpsi(\to l^+ l^-)\KS$        & yes & --                              & $ 1803 \pm 43  $ \\
 $B^0 \rar \psitwos \KS$  & yes          & --                              & $ 142  \pm 13  $ \\
 $B^0 \rar \eta^\prime\KS$               & yes & --                              & $ 37   \pm 16  $ \\
 $B^0 \rar a_1^{\pm} \pi^{\mp}$          & no  & $(39.7 \pm 3.7) \times 10^{-6}$ & $ 7.3  \pm 0.7 $ \\
 $B^0 \rar \D^{*-}(\to D\pi) \pip $      & no  & $(2.57 \pm 0.10)\times 10^{-3}$ & $ 43.8 \pm 2.5 $ \\
 $B^0 \rar \D^- h^+ \ \ \text{;}\ \ B^0 \rar \D^-\mu^+\nu_{\mu}$
                                         & no  & $(2.94 \pm 0.19)\times 10^{-3}$ & $ 281  \pm 20  $ \\
 $B^0 \rar \D^{*-} \rho^+$               & no  & $(14.2 \pm 1.4) \times 10^{-3}$ & $ 34.5 \pm 4.6 $ \\
\hline
$B^0 \rar \{\text{neutral generic decays}\}$ 	& no  & --               & $114\pm 7$  \\
$B^+ \rar \{\text{charged generic decays}\}$ 	& no  & --               & $282\pm 11$ \\
\hline
\end{tabular*}

\vspace{-0.2cm}
\caption{ \label{tab:bbackground}
	\em Summary of \B background modes taken into account for the
	likelihood model. When the yield is varied in the fit, the quoted number of events
	corresponds to the fit results. Otherwise the expected number, taking into account the
	branching ratios and efficiency, is given.}
\end{center}
\end{table*}

We use MC simulated events to study the background from other $B$ 
decays. More than fifty channels were considered in 
preliminary studies, of which twenty are  included
in the final likelihood model -- decays with at least two events expected after
selection.
These exclusive \B background modes are grouped into ten 
different classes that gather decays with similar kinematic and topological
properties: nine for neutral \B decays, one of which accounts for inclusive decays,
and one for charged inclusive \B decays.

Table \ref{tab:bbackground} summarizes the ten background classes that are
used in the fit. When the yield of a class is varied in the Maximum Likelihood fit
the quoted number of events corresponds to the fit results.
For the other modes, the expected number of selected events is
computed by multiplying the selection efficiency (estimated using MC
simulated decays) by the branching fraction, scaled to the dataset 
luminosity ($347\;\mathrm{fb}^{-1}$). The world average branching 
ratios~\cite{hfag,pdg2006} have been
used for the experimentally known decay modes.

\subsection{THE MAXIMUM-LIKELIHOOD FIT}
\label{subsec:ML}

We perform an unbinned extended maximum-likelihood fit to extract
the inclusive $\BztoKspipi$ event yield and the resonant amplitudes.  
The fit uses the variables $\mes$, $\de$, the NN output and the 
SDP to discriminate signal from background. The 
$\dt$ measurement allows the determination of mixing-induced \CP violation
and provides additional continuum background rejection. 

The selected on-resonance data sample is assumed to consist of signal,
continuum background and \B background components, separated by the 
flavor and tagging category of the tag side \B decay. 
The signal likelihood consists of the sum of a correctly 
reconstructed (``truth-matched'', TM) component and a misreconstructed 
(``self-cross-feed'', SCF) component.

The probability density function (PDF) ${\cal P}_i^\cat$ for an
event $i$ in tagging category $\cat$ is the sum of the probability densities 
of all components, namely
\begin{eqnarray}
\label{eq:theLikelihood}
        {\cal P}_i^\cat
        &\equiv& 
                N_{\rm sig} f^\cat_{\rm sig}
                \left[  (1-\fscfave^\cat){\cal P}_{{\rm sig}-\TM,i}^\cat +
                        \fscfave^\cat{\cal P}_{{\rm sig}-\SCF,i}^\cat 
                \right] 
                \nonumber\\[0.3cm]
        &&
                +\; N^\cat_{q\bar q}\frac{1}{2}
                \left(1 + \Qtagi\Atagqq\right){\cal P}_{q\bar q,i}^\cat
                \nonumber \\[0.3cm]
        &&
                +\; \sum_{j=1}^{N^{B^+}_{\rm class}}
                N_{B^+j} f^\cat_{B^+j}
                \frac{1}{2}\left(1 + \Qtagi \Atagj\right){\cal P}_{B^+,ij}^\cat
                \nonumber \\[0.3cm]
        &&
                +\; \sum_{j=1}^{N^{B^0}_{\rm class}}
                N_{B^0j} f^\cat_{B^0j}
                {\cal P}_{B^0,ij}^\cat~,
\end{eqnarray}
where
        $N_{\rm sig}$ is the total number of $\Kspipi$ signal events 
        in the data sample;
        $f^\cat_{\rm sig}$ is the fraction of signal events that are 
        tagged in category $\cat$;
        $\fscfave^\cat$ is the fraction of SCF events in tagging category $\cat$, 
        averaged over the DP;
        ${\cal P}_{{\rm sig}-\TM,i}^\cat$ and ${\cal P}_{{\rm sig}-\SCF,i}^\cat$
        are the products of PDFs of the discriminating variables used
        in tagging category $\cat$ for TM and SCF
        events, respectively; 
        $N^\cat_{q\bar q}$ is the number of continuum events that are 
        tagged in category $\cat$;
        $\Qtagi$ is the tag flavor of the event, defined to be 
        $+1$ for a $\Bz_{\rm tag}$ and $-1$ for a $\Bzb_{\rm tag}$; 
        $\Atagqq$ parameterizes possible tag asymmetry in continuum events; 
        ${\cal P}_{q\bar q,i}^\cat$ is the continuum PDF for tagging 
        category $\cat$;
        $N^{B^+}_{\rm class}$ ($N^{B^0}_{\rm class}$) is the number of 
        charged (neutral) $B$-related background classes considered in the fit,
        namely one (nine);
        $N_{B^+j}$ ($N_{B^0j}$) is the number of expected events in
        the charged (neutral) $B$ background class $j$;
        $f^\cat_{B^+j}$ ($f^\cat_{B^0j}$) is the fraction of 
        charged (neutral) $B$ background events of class $j$
        that are tagged in category $\cat$;
        $\Atagj$ describes a possible tag asymmetry in the charged $B$ background
        class $j$;     
        ${\cal P}_{B^+,ij}^\cat$ is the $B^+$ background PDF for tagging 
        category $\cat$ and class $j$;
        and ${\cal P}_{B^0,ij}^\cat$ is the neutral $B$ background 
        PDF for tagging category $\cat$ and class $j$.
Correlations between the tag and the position in the DP 
 are absorbed in tag-flavor-dependent 
        SDP PDFs that are used for charged \B and continuum
        background.
The PDFs ${\cal P}_{X}^{\cat}$ ($X=\{{\rm sig}\!-\!\TM,\  {\rm sig}\!-\!\SCF,\  q\bar q,\  B^+,\  B^0$)
are the product of the four PDFs of the discriminating variables~\footnote{
Not all the PDFs depend on the tagging category.
The general notations $P_{X,i(j)}^\cat$ and ${\cal P}_{X,i(j)}^{\cat}$ are used for simplicity.},
$x_1 = \mes$, $x_2 = \de$, $x_3 = {\rm NN~output}$ and the triplet
$x_4 = \{\mprime, \thetaprime, \deltat\}$:
\begin{equation}
\label{eq:likVars}
        {\cal P}_{X,i(j)}^{\cat} \;\equiv\; 
        \prod_{k=1}^4 P_{X,i(j)}^\cat(x_k)~,
\end{equation}
where $i$ is the event index and $j$ is a $B$ background class.  The extended likelihood
over all tagging categories is given by
\begin{equation}
        {\cal L} \;\equiv\;  
        \prod_{\cat=1}^{7} e^{-\overline N^\cat}\,
        \prod_{i}^{N^\cat} {\cal P}_{i}^\cat~,
\end{equation}
where $\overline N^\cat$ is the total number of events expected in category 
$\cat$. 

A total of $75$ parameters are varied in the fit. They include the $12$ inclusive yields (signal, four \B background classes and seven continuum yields, one per tagging category) and $30$ parameters for the complex amplitudes from Equation~(\ref{eq:dt}). Most of the
parameters describing the continuum distributions are free in the fit.

\subsubsection{\boldmath THE $\dt$ AND DALITZ PLOT PDFS}
\label{sec:deltaT}

        The SDP PDFs require as input the DP dependent 
        relative selection efficiency, $\varepsilon=\varepsilon(\mprime,\thetaprime)$, 
        and SCF fraction, $\fscf=\fscf(\mprime,\thetaprime)$.
        Both quantities are taken from MC simulation. 
        Away from the DP corners the efficiency is uniform. It 
        decreases when approaching the corners, where one of the 
        three particles in the final state is close to rest so that the 
        acceptance requirements on the particle reconstruction become 
        restrictive.
        Combinatorial backgrounds and hence SCF fractions are large in
        the corners
        of the DP due to the presence of soft tracks.

        For an event~$i$, we define the time-dependent SDP PDFs
        \begin{eqnarray}
                P_{{\rm sig}-\TM,i}(\mprime, \thetaprime, \deltat) &=&
                \varepsilon_i\,(1 - \fscfi)\,\detJi\,\AmpAll~,
                \\[0.3cm]
                P_{{\rm sig}-\SCF,\,i}(\mprime, \thetaprime, \deltat) &=&
                \varepsilon_i\,\fscfi\,\detJi\,\AmpAll~,
        \end{eqnarray}   
        where $P_{{\rm sig}-\TM,i}(\mprime, \thetaprime, \deltat)$
        and $P_{{\rm sig}-\SCF,\,i}(\mprime, \thetaprime, \deltat)$ are normalized. The 
        phase space integration involves the expectation values        
        $\langle \varepsilon\,(1-\fscf)\,\detJ \,F_\kappa F^*_\sigma\rangle$
        and 
        $\langle \varepsilon\,\fscf\,\detJ\, F_\kappa F^*_\sigma\rangle$
        for TM and SCF events, where the indices $\kappa$, $\sigma$ 
        run over all resonances belonging to the signal model.
        The expectation values are model-dependent and are 
        computed with the use of MC integration over the SDP:
        \begin{equation}
        \label{eq:normAverage}
                \langle \varepsilon\,(1-\fscf)\,\detJ\, F_\kappa F^*_\sigma\rangle
                \;=\; \frac{\int_0^1\int_0^1 
                            \varepsilon\,(1-\fscf)\,\detJ\, F_\kappa F^*_\sigma
                        \,d\mprime d\thetaprime}
                       {\int_0^1\int_0^1 \varepsilon\,\detJ\, F_\kappa F^*_\sigma
                        \,d\mprime d\thetaprime}~,
        \end{equation}
        and similarly for 
        $\langle \varepsilon\,\fscf\,\,\detJ\, F_\kappa F^*_\sigma\rangle$,
        where all quantities in the integrands are DP dependent.

        Equation~(\ref{eq:theLikelihood}) invokes the phase 
        space-averaged SCF fraction 
        $\fscfave\equiv\langle\fscf\,\detJ\, F_\kappa F^*_\sigma\rangle$. 
        The PDF normalization  is decay-dynamics-dependent
        and is computed iteratively. We 
        determine the average SCF fractions separately for each tagging category 
        from MC simulation. 
        
        The width of the dominant 
        resonances is large compared 
        to the mass resolution for TM events (about $8\mevcc$ core Gaussian
        resolution). We  therefore neglect resolution effects in the TM 
        model.  
        Misreconstructed events have a poor mass resolution that strongly 
        varies across the DP. It is described in the fit by a 
        $2\times 2$-dimensional resolution function
        \begin{equation}
        \label{eq:rscf}
                \Rscf(\mprime_r,\thetaprime_r,\mprime_t,\thetaprime_t)~,
        \end{equation}
        which represents the probability to reconstruct at the coordinate
        $(\mprime_r,\thetaprime_r)$ an event that has the true coordinate 
        $(\mprime_t,\thetaprime_t)$. It obeys the unitarity condition
        \begin{equation}
                \intl_0^1\intl_0^1 
                \Rscf(\mprime_r,\thetaprime_r,\mprime_t,\thetaprime_t)
                \,d\mprime_r d\thetaprime_r = 1~,
        \end{equation}
        and is convolved with the signal model. 
        The $\Rscf$ function is obtained from MC simulation.

       We use the signal model described in Section~\ref{sec:kinematics}. 
        It contains the dynamical information and is connected with $\dt$ via 
        the matrix element in Equation~(\ref{eq:dt}), which serves as the PDF.
        The PDF is diluted 
        by the effects of mistagging and the limited vertex 
        resolution~\cite{rhopipaper}. 
        The $\deltat$ resolution function for signal (both TM and SCF) and \B background 
        events is a sum of three Gaussian distributions, with parameters 
        determined by a fit to fully reconstructed $\Bz$ 
        decays~\cite{BabarS2b}. 

        The SDP- and $\dt$-dependent PDFs factorize for the 
        charged $B$ background modes, but not necessarily
        for the neutral $B$ background due to $\BzBzb$ mixing.

        The charged \B background
                contribution to the likelihood (Equation~(\ref{eq:theLikelihood}))
                involves 
                the parameter $\Atag$, multiplied by the tag flavor $\Qtag$ of 
                the event. In the presence of significant tag-``charge'' 
                correlation (represented by an effective 
                flavor-tag-versus-Dalitz-coordinate correlation),
                it parameterizes possible direct \CP violation in these events.
                We use distinct SDP PDFs for each 
                reconstructed $B$ flavor tag, and a flavor-tag-averaged PDF for 
                untagged events. The PDFs are obtained from MC simulation and are 
                described by histograms.
                The $\dt$ resolution parameters are determined by a fit to fully 
                reconstructed $\Bp$ decays. 
                For the $\Bp$ background class we adjust the 
                effective lifetime to account for the misreconstruction of the 
                event that modifies the nominal $\dt$ resolution function.

        The neutral $B$ background is parameterized with PDFs that
                depend on the flavor tag of the event. In the case of \CP
                eigenstates, correlations between the flavor tag and the Dalitz 
                coordinate are expected to be small. However, non-\CP  eigenstates,
                such as $a_1^\pm\pi^\mp$, may exhibit such correlations. Both types 
                of decays can have direct
                and mixing-induced \CP  violation. A third type of decay
                involves charged $\D$ mesons and does not exhibit mixing-induced
                \CP  violation, but usually has a strong correlation between the
                flavor tag and the DP coordinate (the $\D$ meson charge), because 
                it consists of $B$-flavor eigenstates. Direct \CP violation is also
                possible in these decays, though it is set to zero in the nominal model.
                The DP PDFs are obtained from MC simulation and are 
                described by histograms.
                For neutral $B$ background, the signal $\dt$ resolution model 
		is assumed. 

        The DP
                treatment of the continuum events is similar to that used
                for charged $B$ background. 
                The SDP PDF for continuum background is 
                obtained from on-resonance events selected in the
                $\mes$ sidebands and corrected for feed-through
                from \B decays. A large number of cross checks have been 
                performed to ensure the high fidelity of the empirical shape 
                used. Analytical models were found to be insufficient.
                The continuum $\deltat$ distribution is parameterized as the sum of 
                three Gaussian distributions with common mean and 
                three distinct widths that scale the $\dt$ per-event error. 
                This yields six shape parameters that are determined by 
                the fit.
                The model is motivated by the observation that 
                the $\dt$ average is independent of its error, and that the 
                $\dt$ RMS depends linearly on the $\dt$ error.

\subsubsection{DESCRIPTION OF THE OTHER VARIABLES}
\label{sec:likmESanddE}

        The $\mes$ distribution of TM signal events is
                parameterized by a bifurcated Crystal Ball 
                function~\cite{PDFsCB,Oreglia:1980cs,Gaiser:1982yw},
                which is a combination of a one-sided Gaussian and 
                a Crystal Ball function. The mean and two widths of this function
                are determined by the fit. A non-parametric
                function is used to describe the SCF signal component.
	        The $\de$ distribution of TM events is
                parameterized by a double Gaussian function.
                Misreconstructed events are described by 
                a non-parametric function.
                
        Both $\mes$ and $\de$ PDFs are described by non-parametric
                functions for all $B$ background classes. Exceptions to
                this are the $\mes$ PDFs for $B^0 \to D^-\pi^+$ and 
                $B^0 \to J/ \psi K^0_S$ components, and the $\Delta E$
                PDF for $B^0 \to D^-\pi^+$, which are the same as the corresponding
                distributions of TM signal events.
                
        The $\mes$ and $\de$ PDFs for continuum events are
                parameterized with an ARGUS shape function~\cite{argus} and 
                a first-order polynomial, respectively, with parameters 
                determined by the fit.

        We use non-parametric functions to empirically describe the distributions 
                of the NN output
                found in the MC simulation for TM and SCF signal events, 
                and for \B background events. We distinguish tagging categories 
                for TM signal events to account for differences observed in the 
                shapes.

        The continuum NN distribution is parameterized by a third-order 
        polynomial that is constrained to take positive values in the 
        range populated by the data. 
        The coefficients of the polynomial are determined by the fit.
        Continuum events exhibit a correlation between the DP 
        coordinate
        and the shape of the event that is exploited in the NN. 
        To correct for residual effects,
        we introduce a linear dependence of the polynomial coefficients
        on the distance of the DP coordinate from the kinematic 
        boundaries of the DP. The parameters describing this
        dependence are determined by the fit.

\section{FIT RESULTS}
\label{sec:fitResults}

\begin{figure}[htbp]
  \centerline{  \epsfxsize7cm\epsffile{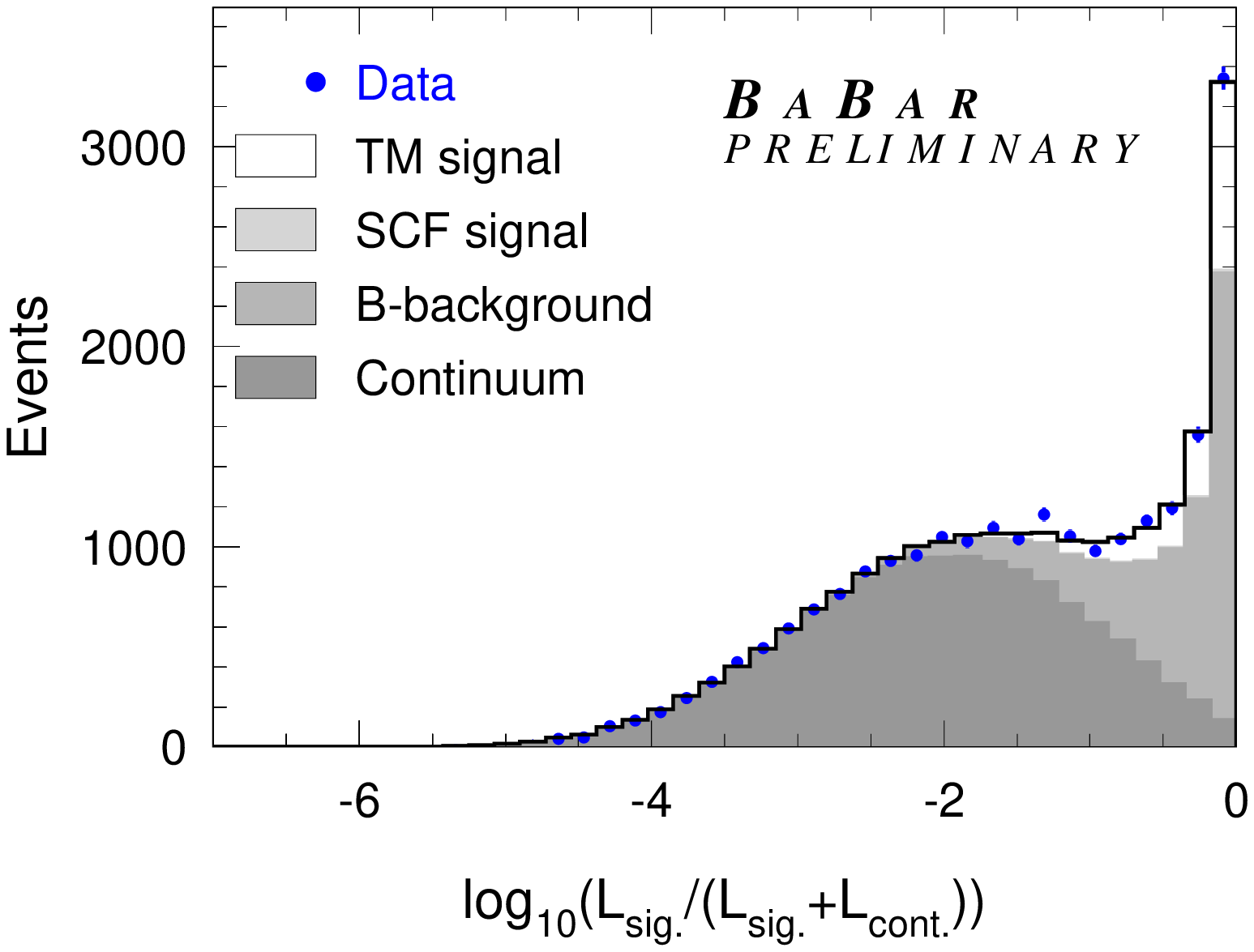}
 \epsfxsize7cm\epsffile{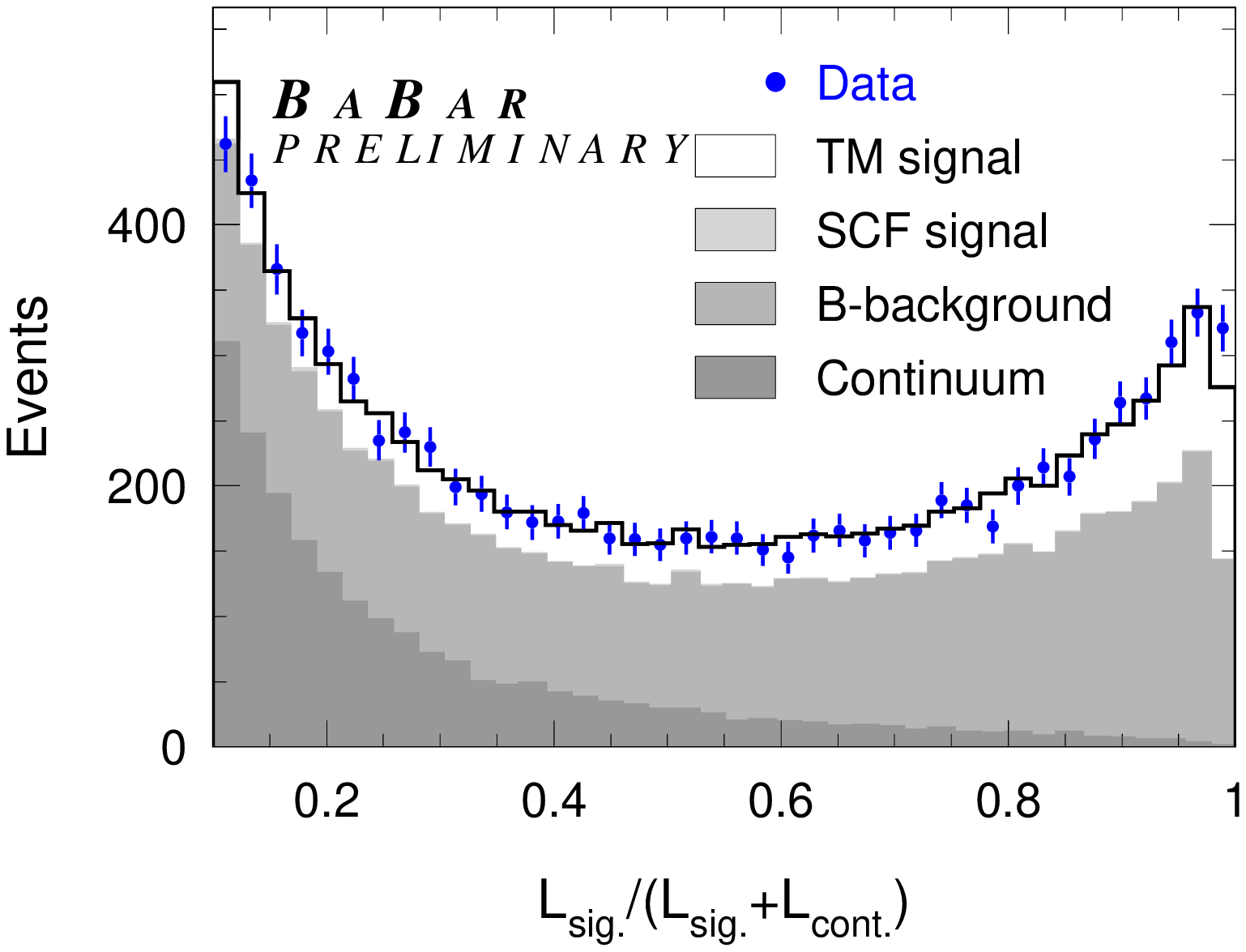}}
  \caption{\em \label{fig:lkl} Distributions of the likelihood ratio (signal/background)
         for all events entering the fit (left)
        and in the signal-like region (right). The dots with error bars give 
        the on-resonance data. The solid histogram shows the
        projection of the fit result. The dark,
        medium and light shaded areas represent respectively the contribution
        from continuum events, the sum of continuum events 
        and the $B$ background expectation, and the sum of these and 
        the misreconstructed signal events. The last contribution is hardly visible due to its
        small fraction. In both distributions the $\D^- \pip$ and $\jpsi \KS$ bands are removed from the DP.}
\end{figure}

\begin{figure}[htbp]
  \centerline{  \epsfxsize7cm\epsffile{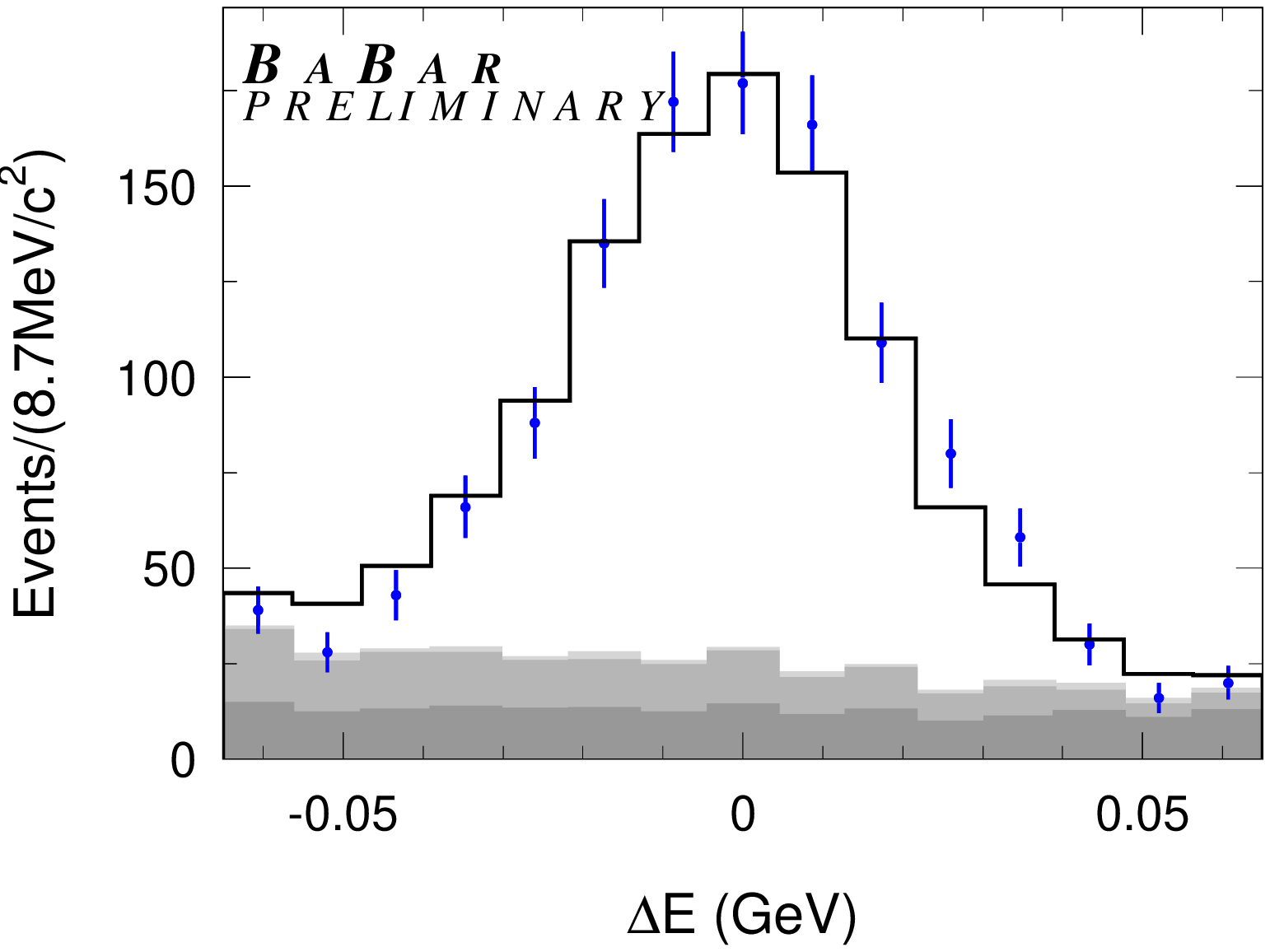}
 \epsfxsize7cm\epsffile{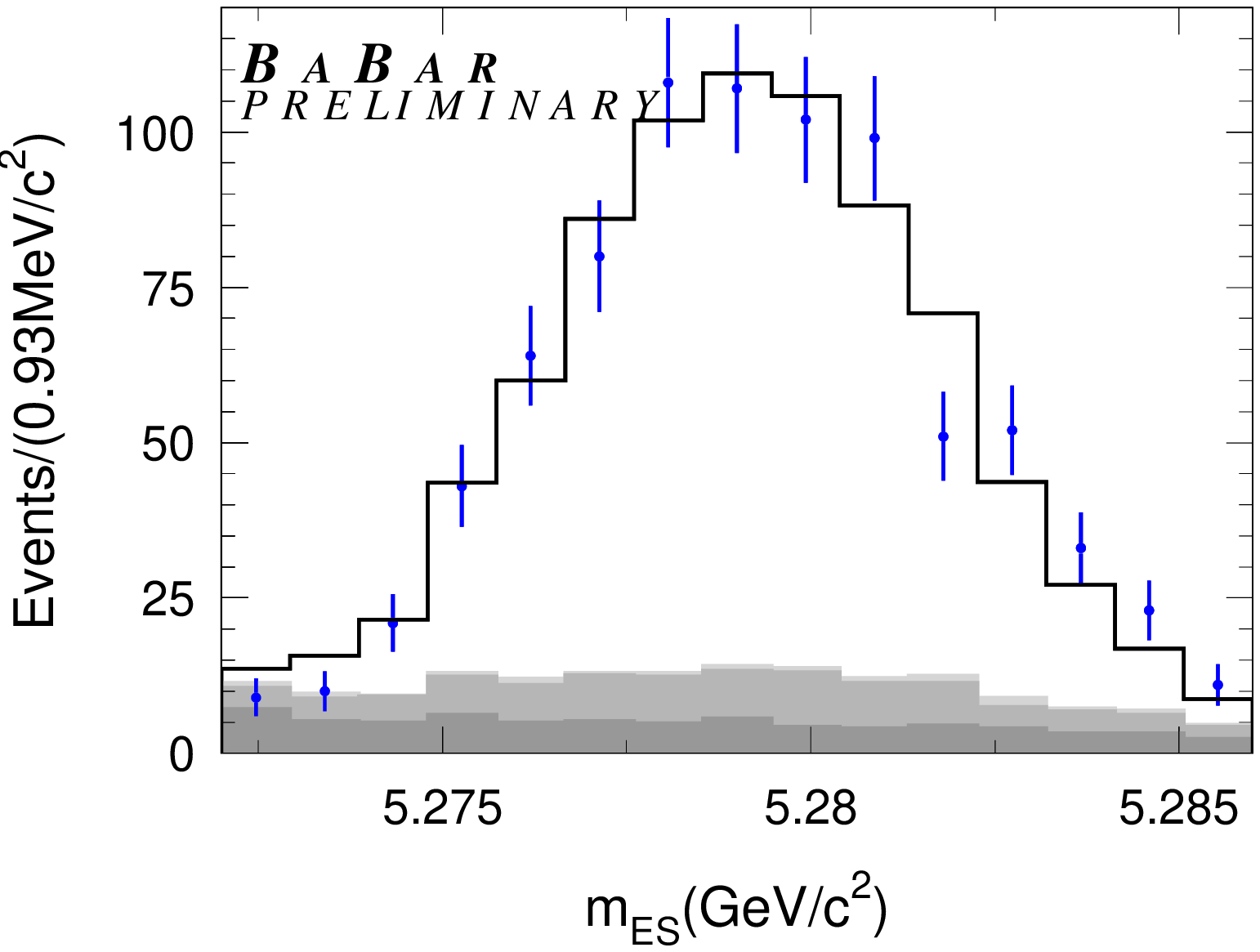}}
  \centerline{  \epsfxsize7cm\epsffile{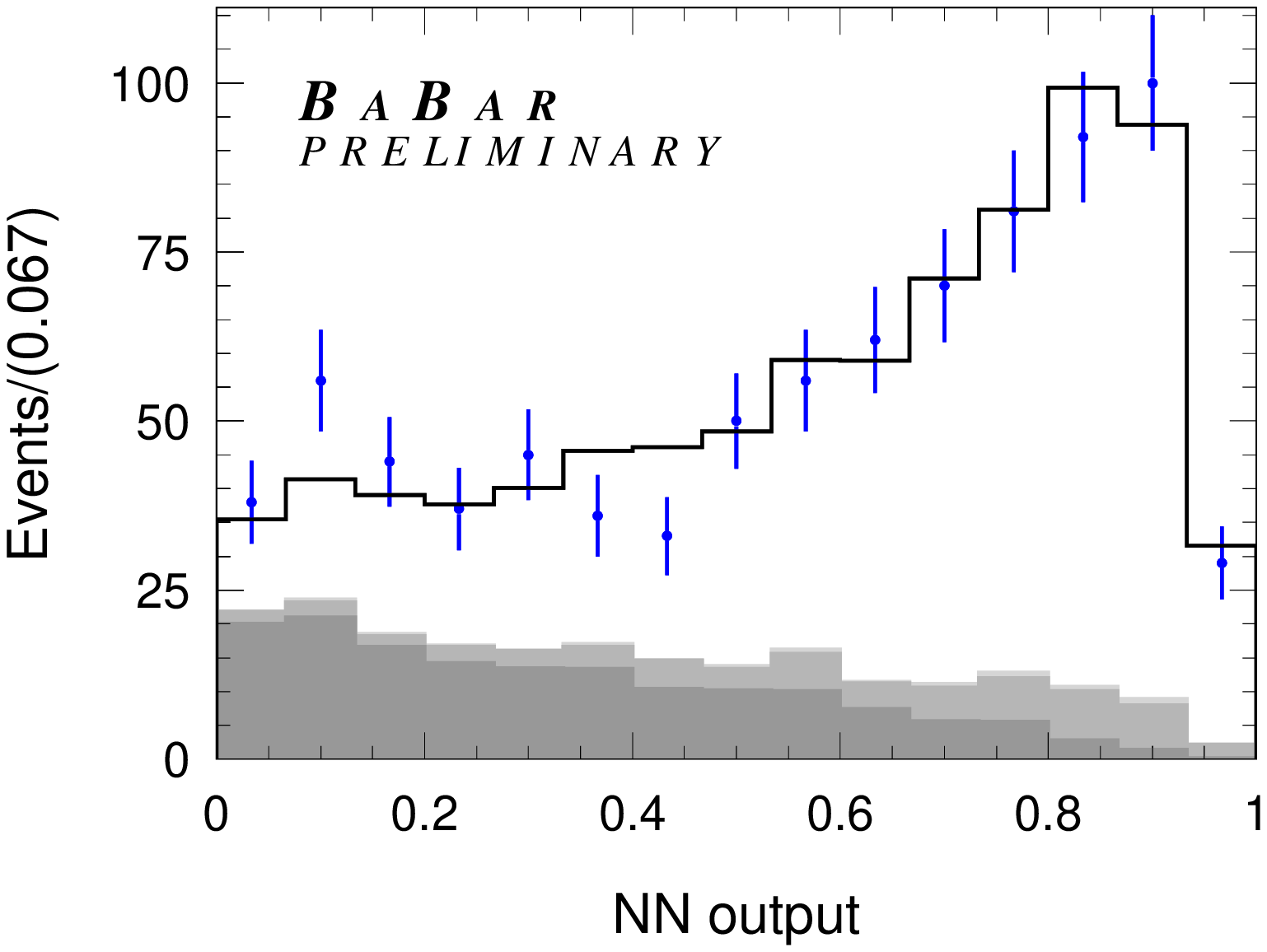}
 \epsfxsize7cm\epsffile{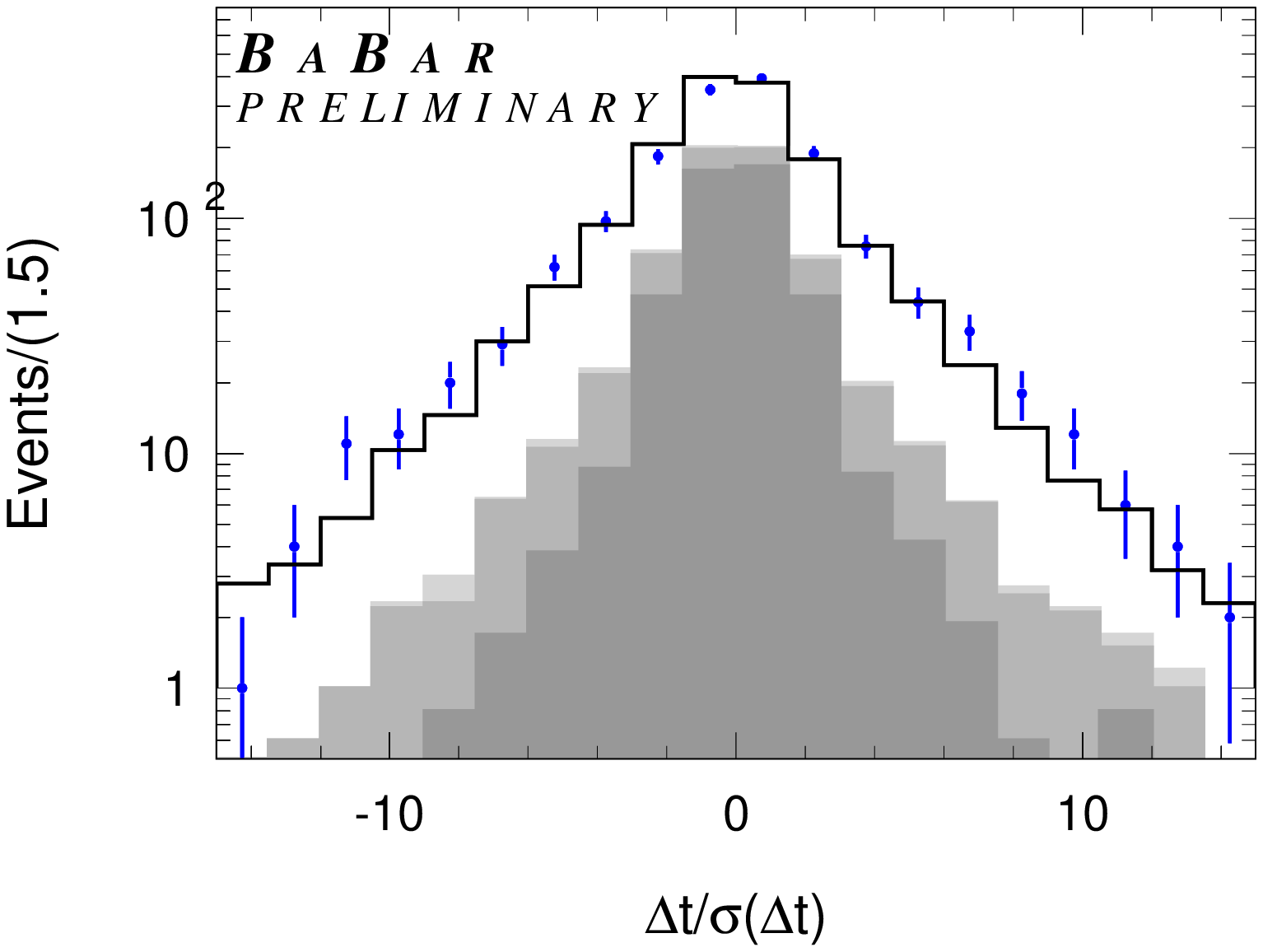}}
  \centerline{  \epsfxsize7cm\epsffile{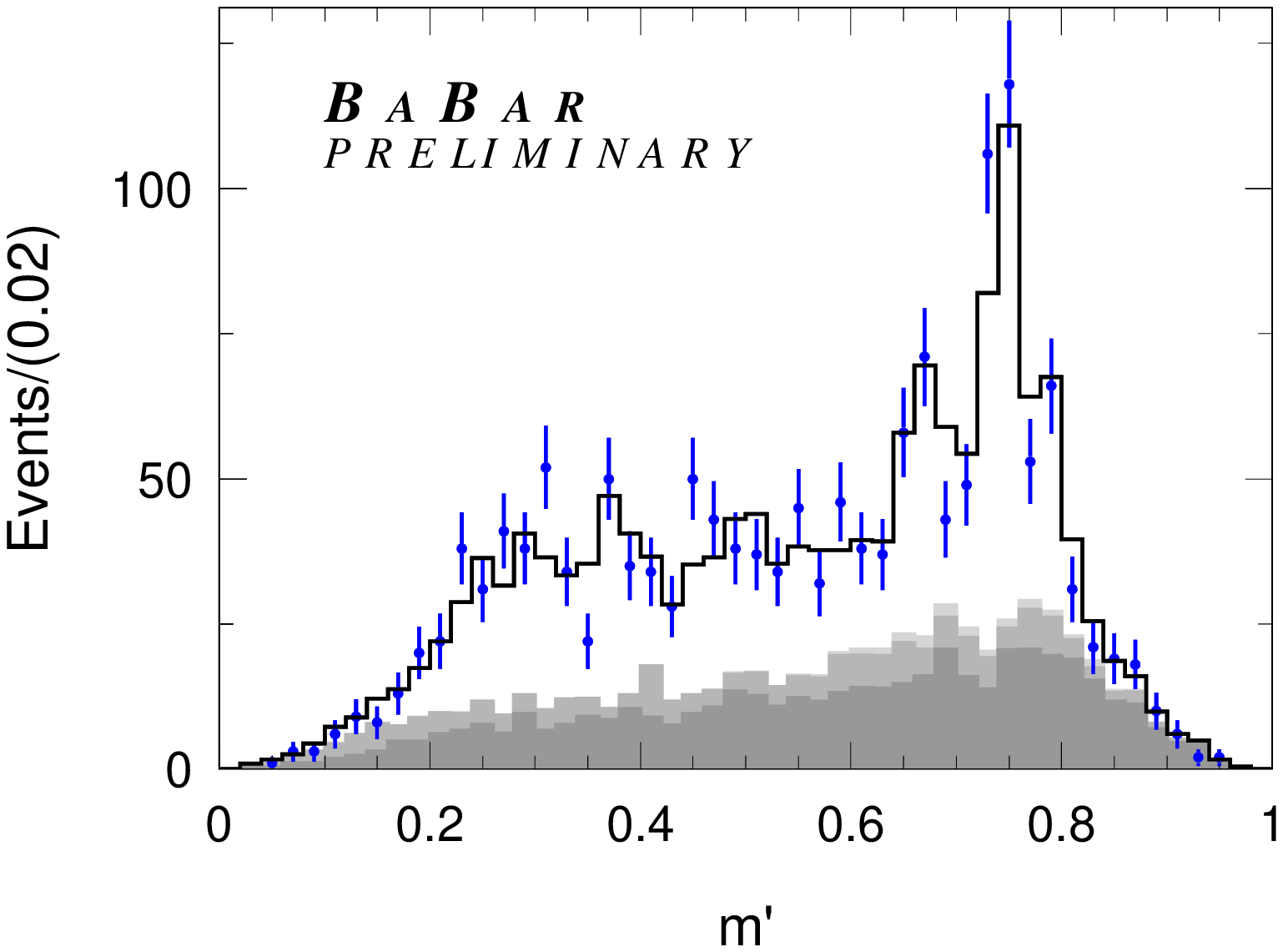}
  \epsfxsize7cm\epsffile{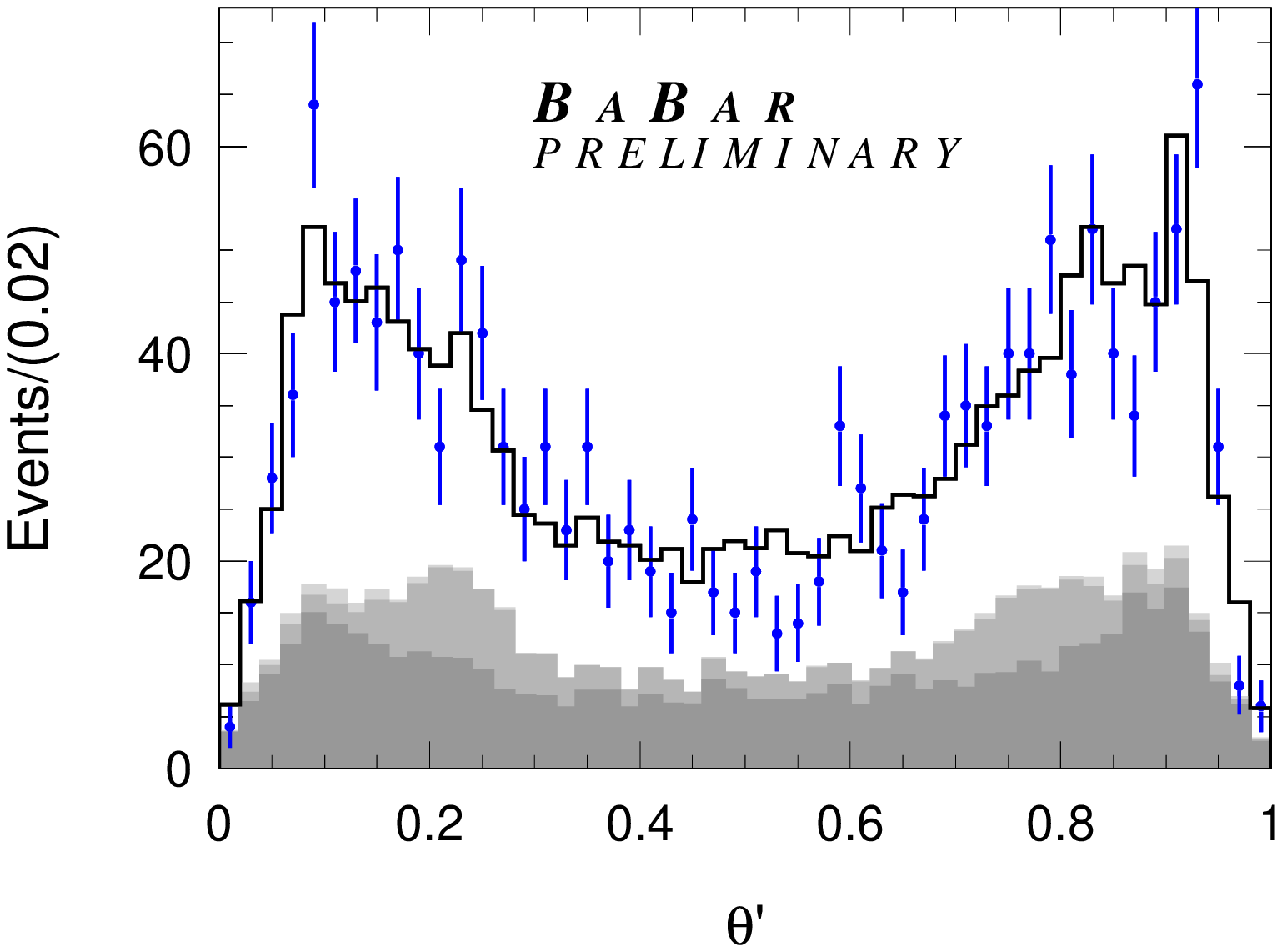}}
  \caption{\em \label{fig:projections}     Distributions of (top to bottom, left to right) $\de$, $\mes$, 
        NN output, $\dt/\sigma(\dt)$,  $\mprime$ and $\thetaprime$ for samples 
        enhanced in $\BztoKspipi$ signal. The dots with error bars give 
        the on-resonance data. The solid histogram shows the
        projection of the fit result. The dark,
        medium and light shaded areas represent respectively the contribution
        from continuum events, the sum of continuum events 
        and the $B$ background expectation, and the sum of these and 
        the misreconstructed signal events. The last contribution is hardly visible due to its
        small fraction. In all these distributions the $\D^- \pip$ and $\jpsi \KS$ bands are removed from the DP.}
\end{figure}

\begin{figure}[htbp]
  \centerline{  \epsfxsize8.5cm\epsffile{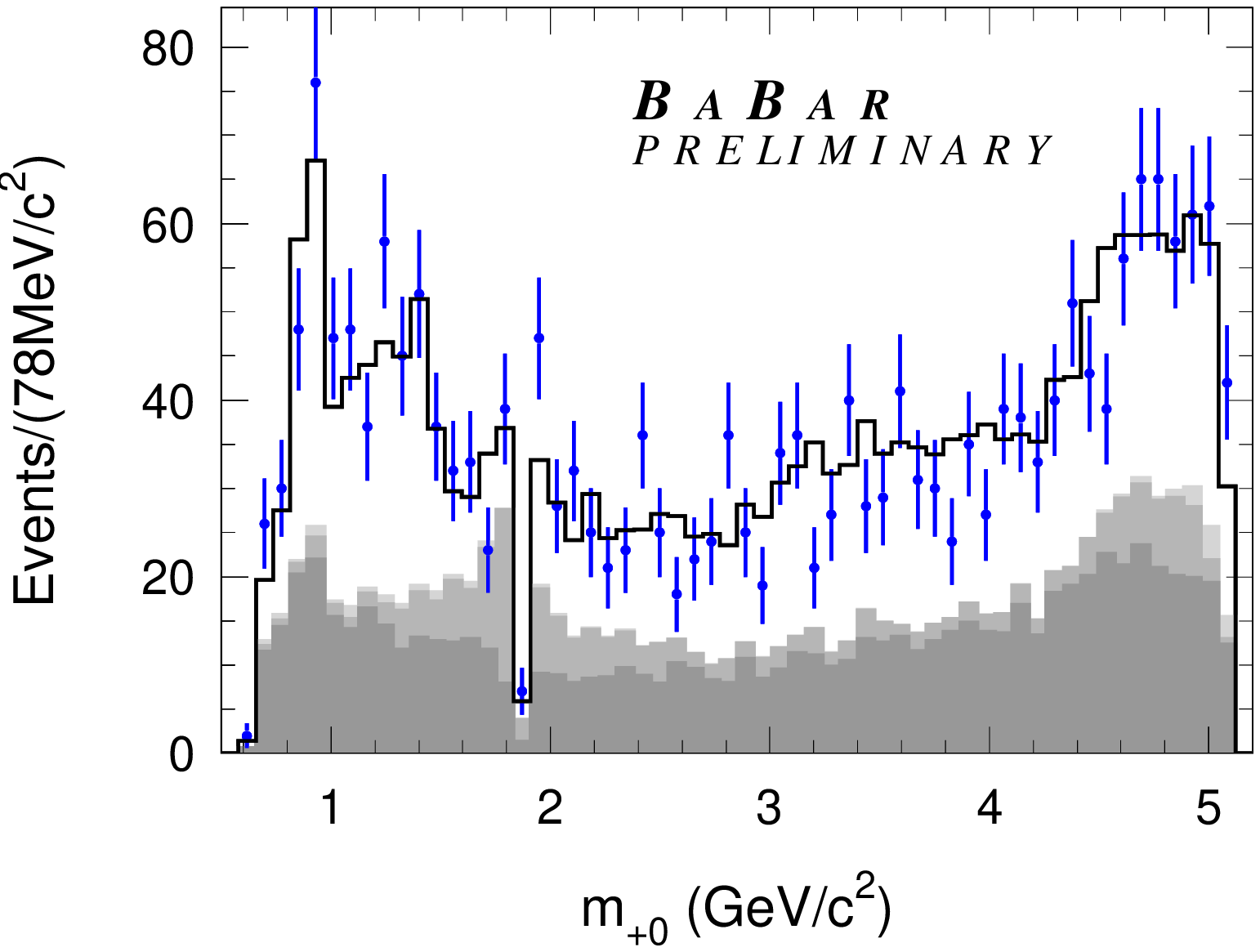}
	\epsfxsize8.5cm\epsffile{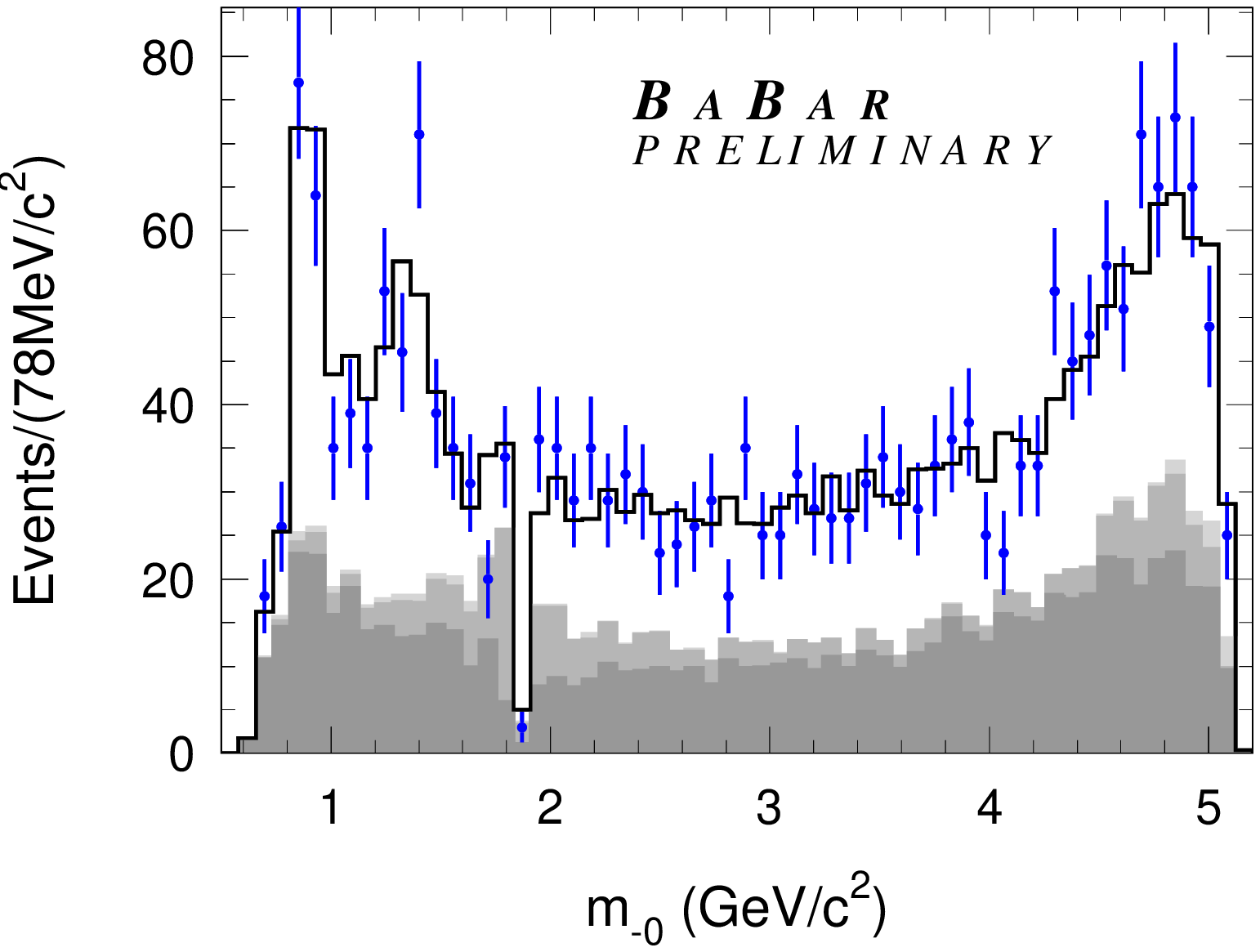}}	
 \centerline{	\epsfxsize8.5cm\epsffile{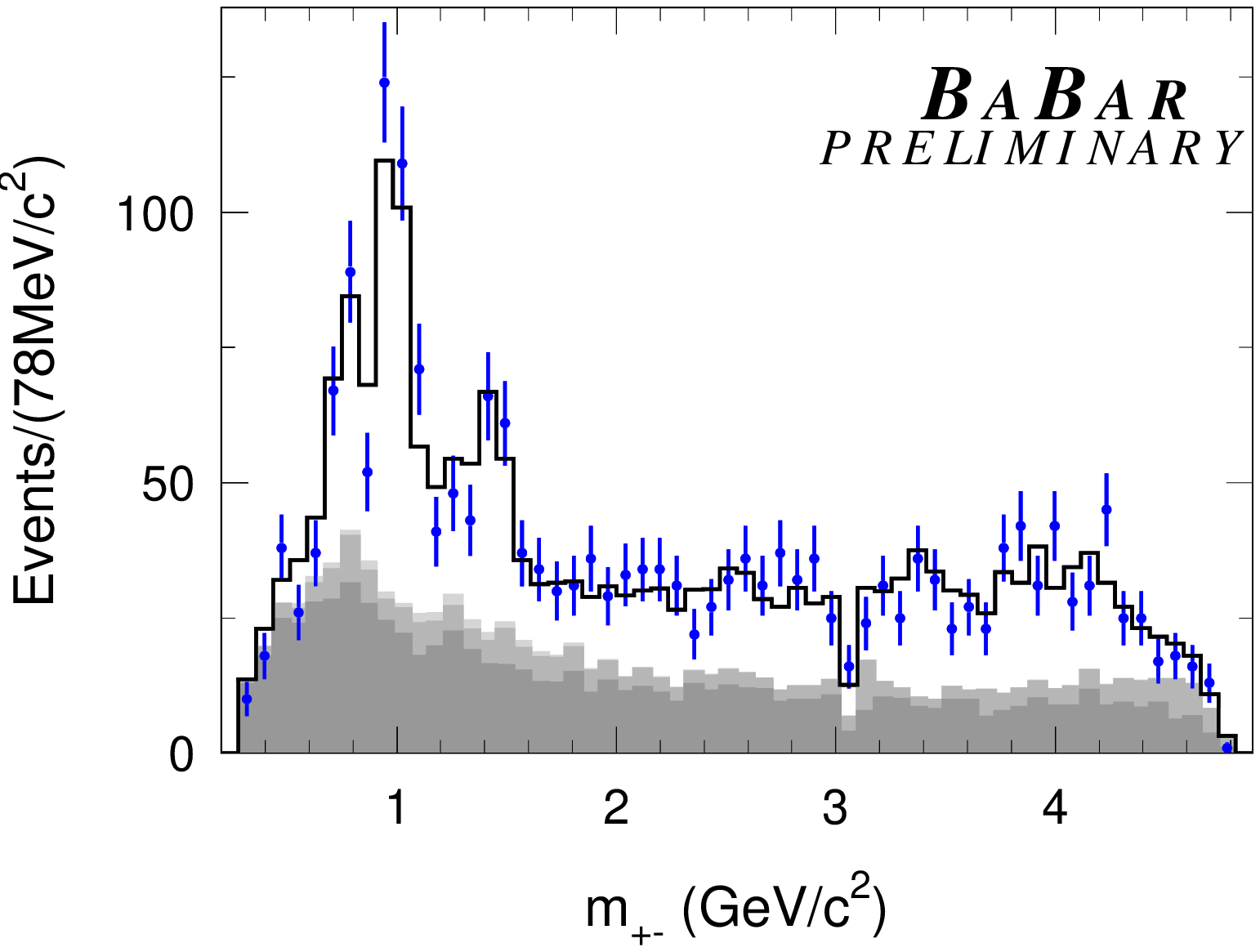}}
  \caption{\em \label{fig:projmasses} Distributions of $m(K_s^0 \pi^+)$ (top left), 
	$m(K_s^0 \pi^-)$ (top right), and $m(\pi^+\pi^-)$ (bottom) for samples 
        enhanced in $\BztoKspipi$ signal. The solid histogram shows the
        projection of the fit result. The dark,
        medium and light shaded areas represent respectively the contribution
        from continuum events, the sum of continuum events 
        and the $B$ background expectation, and the sum of these and 
        the misreconstructed signal events. The last contribution is hardly visible due to its
        small fraction. In all these distributions the $\D^- \pip$ and $\jpsi \KS$ bands are removed from the DP.}
\end{figure}

\begin{figure}[htbp]
  \centerline{  \epsfxsize8.5cm\epsffile{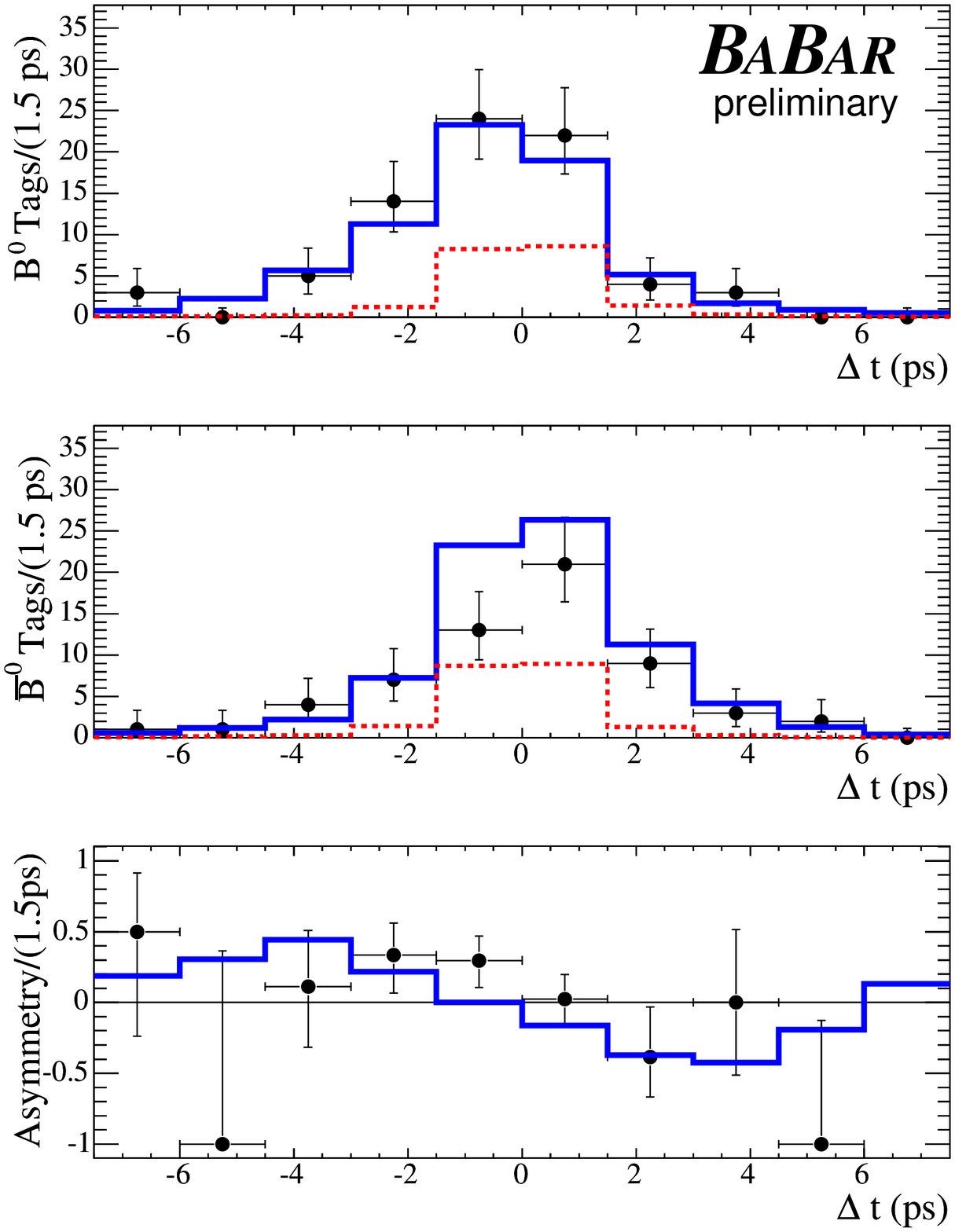}
	\epsfxsize8.5cm\epsffile{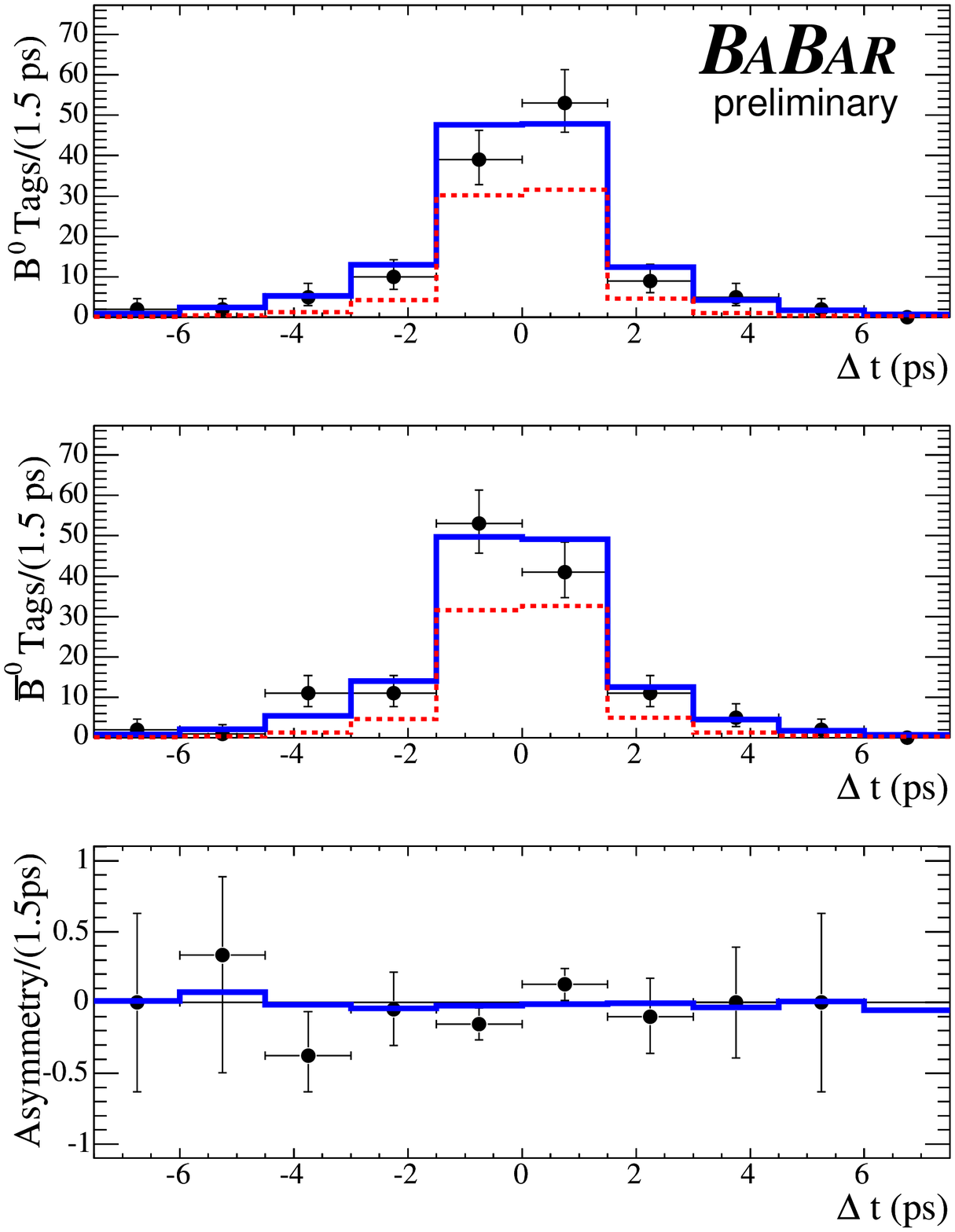}}
  \caption{\em \label{fig:dtAssym} Distributions of $\dt$ when the $\Bz_{\rm tag}$ is a $\Bz$ (top), $\Bzb$ (middle) and the derived $\dt$ asymmetry (bottom). Plots on the left (right) hand side, correspond to events in the $\fI \KS$ ($\rhoI \KS$) region. The solid line is the total PDF, the dashed line is the continuum only PDF and points with error bars represent data. These distributions correspond to samples enhanced in $\BztoKspipi$ signal, where the $\D^- \pip$ and $\jpsi \KS$ bands are removed from the DP.}
\end{figure}

The maximum-likelihood fit of $22525$ candidates results in a $\BztoKspipi$ event yield of
$2172\pm 70$ and a continuum yield of $14272\pm 126$, where the errors are statistical only.
Figure~\ref{fig:lkl} shows distributions of the likelihood ratio (signal/background) for all
the events entering the fit and for the signal-like region. Figure~\ref{fig:projections} shows
distributions of $\de$, $\mes$, NN output, $\dt/\sigma(\dt)$, where $\sigma(\dt)$
is the per-event error on $\dt$, as well as the DP 
variables $\mprime$ and $\thetaprime$,
which are enhanced in signal content by requirements on the signal-to-continuum 
likelihood ratios of the other discriminating variables. Figure~\ref{fig:projmasses}
shows similar distributions for $m(K_s^0 \pi^+)$,  $m(K_s^0 \pi^-)$ and $m(\pi^+\pi^-)$.
These distributions indicate the good quality of the fit.
Signal enhanced distributions of $\dt$ and $\dt$ asymmetry for events in the regions of $\fI \KS$ and $\rhoI \KS$
are shown in Figure~\ref{fig:dtAssym}.

In the fit, we measure directly the relative magnitudes and phases 
of the different components of the signal model. The results
are given together with their statistical
errors in Table~\ref{tab:resultsAmps}.
The measured relative amplitudes $c_\sigma$, where the index represents an intermediate resonance,
are used to extract the  Q2B parameters, for which the definitions are given below.

\begin{table}[hbtp]
\begin{center}
\caption{\em
Summary of fit results for the magnitudes $|c_\sigma|$ and phases $\phi$ in degrees of the resonant amplitudes. 
The quoted error is statistical only.}
\label{tab:resultsAmps}
\begin{tabular}{ccccc}
\hline
Resonance Name           & $|c_\sigma|$    &  $\phi[\rm degrees]$     &   $|\overline{c}_\sigma|$ ($|\overline{c}_{\overline{\sigma}}|$)&   $\overline{\phi}[\rm degrees]$  \\
\hline\hline
$\fI   \KS$ 		& $  4.0\ph{0}    $  & $  0.0\ph{0}      $  & $2.8 \pm 0.7 $  & $-88.6 \pm 21.3$  \\
$\rhoI \KS$ 		& $0.10 \pm 0.02  $  & $58.6 \pm 16.4  $  & $0.09 \pm 0.02   $ & $21.3 \pm 21.2$  \\
$f_0(1300)\KS$ 		& $1.9 \pm 0.4    $  & $117.6  \pm 22.6$  & $1.1 \pm 0.3     $ & $-15.2 \pm 23.8$ \\
Nonresonant   		& $3.0 \pm 0.6    $  & $13.8 \pm 14.3  $  & $3.7 \pm 0.5     $ & $-16.2 \pm 17.3$ \\
$\KstarpI\pi^-$ 	& $0.136 \pm 0.021$  & $-60.7 \pm 18.5 $  & $0.113 \pm 0.018 $ & $102.6 \pm 22.9$ \\
$\KstarpII\pi^-$ 	& $4.9 \pm 0.7    $  & $-82.4 \pm 16.8 $  & $7.1 \pm 0.9     $ & $79.2 \pm 20.5$  \\
$f_2(1270)\KS$ 	        & $0.011 \pm 0.004$  & $62.9 \pm 23.3  $  & $0.010 \pm 0.003$ & $-73.9 \pm 27.8$ \\
$\chi_{c0}(1P)\KS$      & $0.34 \pm 0.15  $  & $68.7 \pm 31.1  $  & $0.40 \pm 0.11   $ & $154.5 \pm 28.6$ \\
\hline
\hline
\end{tabular}

\end{center}
\end{table}

For a resonant decay mode $\sigma$
which is a \CP eigenstate,
the following Q2B parameters are extracted: 
the angle
$2\beta_{\rm eff}$, defined as:
     \begin{equation}
       2\beta_{\rm eff}(\sigma) = arg(c_{\sigma}\overline c_{\sigma}^*)~,
     \end{equation} \label{eq:2betaeff}
and the direct and mixing-induced \CP asymmetries, defined as:
    \begin{equation}
    C(\sigma) = \frac{|c_{\sigma}|^2 - |\overline c_{\sigma}|^2}{|c_{\sigma}|^2 + 
|\overline c_{\sigma}|^2}~, 
    \end{equation} \label{eq:C}
    \begin{equation}
    S(\sigma) = \frac{  2{\cal I}m(\overline c_{\sigma}c_{\sigma}^*)}{|c_{\sigma}|^2 + 
|\overline c_{\sigma}|^2}~;
    \end{equation} \label{eq:S}
for a self-tagging resonant decay mode $\sigma$ such as $\Bz \to \KstarpI \pim$,
the direct \CP asymmetry is defined as:
    \begin{equation}
       A_{\CP}(\sigma) =  \frac{|\overline c_{\overline{\sigma}}|^2 - |c_{\sigma}|^2}
       {|\overline c_{\overline{\sigma}}|^2 + |c_{\sigma}|^2}~;
    \end{equation} \label{eq:Acp}
the relative isobar phase between two resonances, $\sigma$ and $\kappa$ is defined by
    \begin{equation}
      \Delta \phi(\sigma, \kappa) = arg(c_{\sigma} c_{\kappa}^*)~;
    \end{equation}
and the similar quantity for a self-tagging resonant decay mode, $\sigma$ 
and its charge conjugate $\overline{\sigma}$ is:
    \begin{equation}
      \Delta \phi(\sigma, \overline{\sigma}) = arg(c_{\sigma} \overline{c}_{\overline{\sigma}}^*)~;
    \end{equation}
recall that we use a convention in which the $\Bzb$ decay amplitudes have absorbed the phase from $\BzBzb$ mixing.
Finally, we also extract the relative fraction $f$ of a Q2B channel $\sigma$, which is calculated as:
      \begin{equation}
      f(\sigma) = \frac{(|c_{\sigma}|^2 + |\overline c_{\sigma}|^2)\langle F_{\sigma}F^*_{\sigma}\rangle}
{\sum_{\mu\nu}{(c_{\mu}c_{\nu}^* + \overline c_{\mu}\overline c_{\nu}^*)\langle F_{\mu}F^*_{\nu}\rangle}}~,
      \end{equation}
where
      \begin{equation}
         \langle F_{\mu}F^*_{\nu}\rangle = \int{F_{\mu}F^*_{\nu}ds_+ds_-}~.
      \end{equation}

Two approaches were used to extract the statistical uncertainties of Q2B parameters. 
The first approach uses a linear approximation, whereby errors are evaluated assuming
 that the likelihood function is a multivariate Gaussian, which is defined by the fit
 result and covariance matrix. In the second approach we
perform a likelihood scan, fixing the scanned Q2B parameter at several
consecutive values and repeating the fit to the data.  The error on the
Q2B parameter is obtained from the shape of the likelihood function near
the minimum.  Since the Q2B parameters are not directly used in the fit,
we fix instead certain parameters that allow the resulting likelihood
curve to be trivially interpreted in terms of the Q2B parameter of interest.
The likelihood scan approach does not rely on any
assumption about the covariance matrix.  We use this approach wherever
the fit variables that determine a Q2B parameter are found to have
non-Gaussian errors.

\begin{table*}[htbp]
\centering

\caption{\label{tab:resultsQ2B}
\em 
Summary of results for the Q2B parameters. The first quoted error is statistical, the second is 
systematic and the third is DP signal model uncertainty. Parameters for which the statistical error have 
been obtained from a likelihood scan are marked by $\dag$. Phases are in degrees and relative fractions in $\%$.}

\begin{minipage}{\textwidth}
\centering
{\small
\begin{tabular}{cc|cc}
\hline\hline
{\bf Parameter} & {\bf Value} & {\bf Parameter} & {\bf Value} \\
\hline\hline  \\[-0.4cm]
$C(f_0(980)K^0_S)$                     & $0.35 \pm 0.27\pm 0.07 \pm 0.04$ &
$C(\rho^0(770)K^0_S)$                  & $0.02 \pm 0.27\pm 0.08 \pm 0.06$ \\
$^\dag 2\beta_{\rm eff}(f_0(980)K^0_S)$    & $(89^{+22}_{-20}\pm 5 \pm 8)^\circ$ &
$^\dag 2\beta_{\rm eff}(\rho^0(770)K^0_S)$ & $(37^{+19}_{-17}\pm 5 \pm 6)^\circ$ \\
$^\dag S(f_0(980)K^0_S)$               & $-0.94^{+0.07 + 0.05}_{-0.02 -0.03} \pm 0.02$ &
$^\dag S(\rho^0(770)K^0_S)$            & $0.61 ^{+0.22}_{-0.24}\pm 0.09 \pm 0.08$  \\
$f(f_0(980)K^0_S)$                     & $14.3^{+2.8}_{-1.8}\pm 1.5 \pm 0.6$ &
$f(\rho^0(770)K^0_S)$                  & $9.0 \pm 1.4\pm 1.1 \pm 1.1$ \\  \\[-0.4cm] \hline \\[-0.4cm]

$A_{\CP}(\KstarpI\pim)$                 & $-0.18 \pm 0.10 \pm 0.03 \pm 0.03$ &
$^\dag \Delta\phi(f_0 K^0_S,\rho^0 K^0_S)$
                                       & $(-59^{+16}_{-17}\pm 6 \pm 6)^\circ$ \\
$^\dag \Delta\phi(\KstarI\pi)$~\footnote{
Abbreviation for $\Delta\phi(\KstarpI\pim, \KstarmI\pip)$.}
                                       & $(-164 \pm 24 \pm 12 \pm 15)^\circ$ && \\
$f(K^*(892)\pi)$                       & $11.7 \pm 1.3\pm 1.3 \pm 0.6$ && \\  \\[-0.4cm] \hline \\[-0.4cm]

$f(K^{*}(1430)\pi)$                    & $38.9 \pm 2.5 \pm 0.7 \pm 1.3$ &
$f(NR)$                                & $25.6 \pm 2.5\pm 1.9 \pm 0.5$ \\
$f(f_0(1300)K^0_S)$                    & $6.3 \pm 1.3\pm 0.6 \pm 0.3$ &
$f(f_2(1270)K^0_S)$                    & $2.1 \pm 0.8\pm 0.0 \pm 0.2$ \\
$f(\chi_{c0}(1P)K^0_S)$                & $1.2 \pm 0.5\pm 0.0 \pm 0.1$ &\\ \\[-0.4cm]
\hline \hline
\end{tabular} 
}
\end{minipage}
\end{table*}

\begin{figure}[htbp]
  \centerline{ 
  \epsfxsize9cm\epsffile{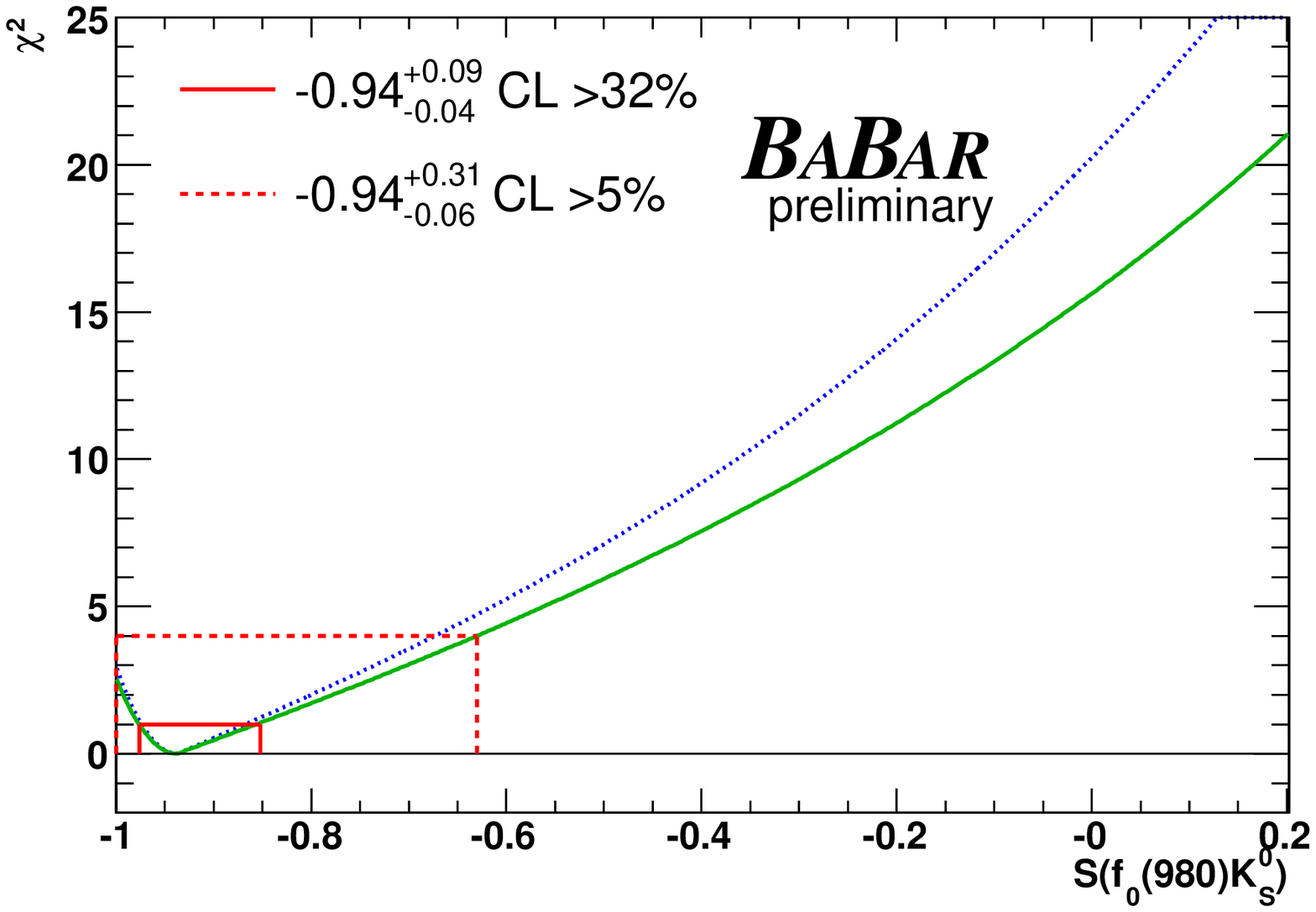}
 \epsfxsize9cm\epsffile{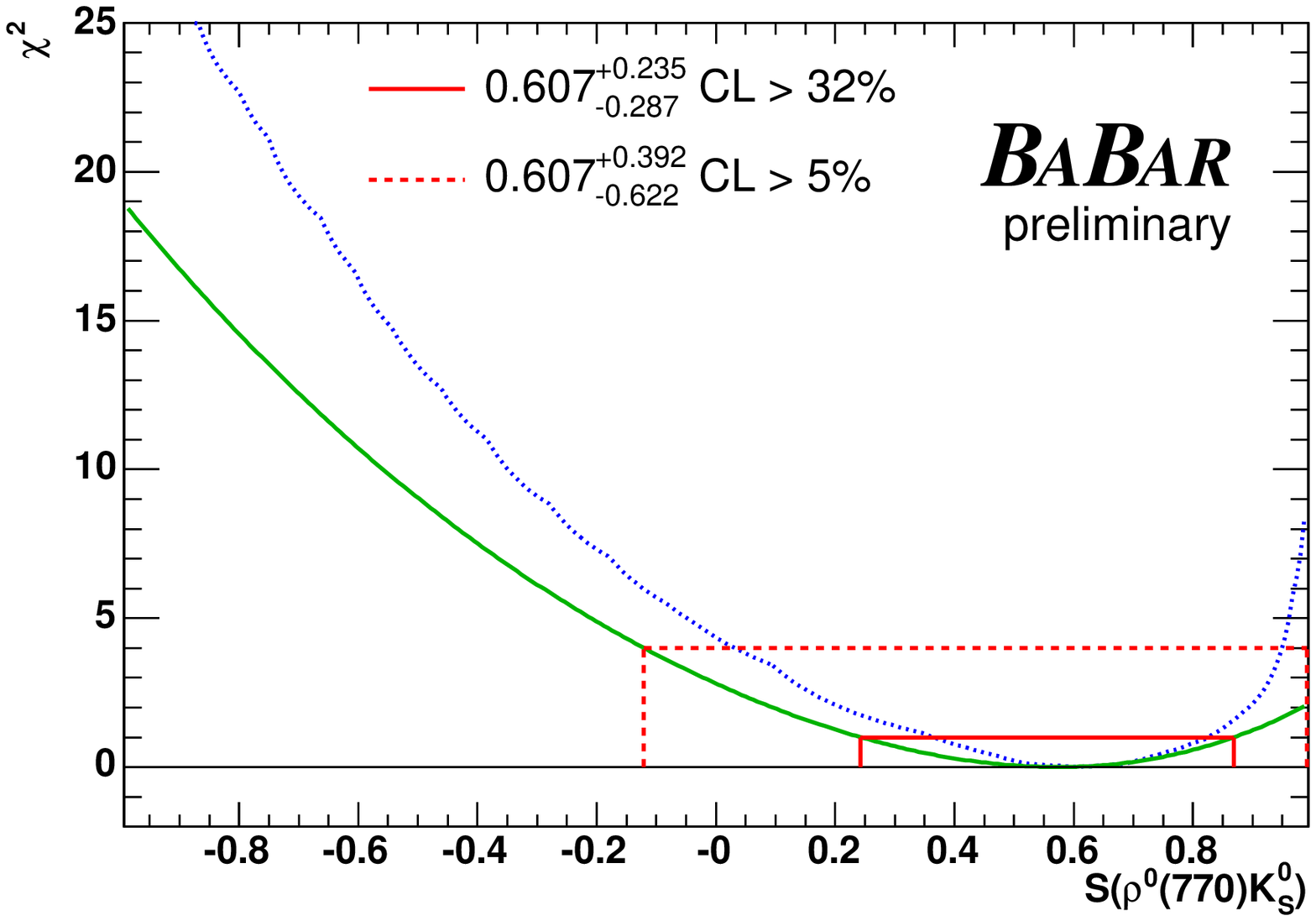}}
  \centerline{  \epsfxsize9cm\epsffile{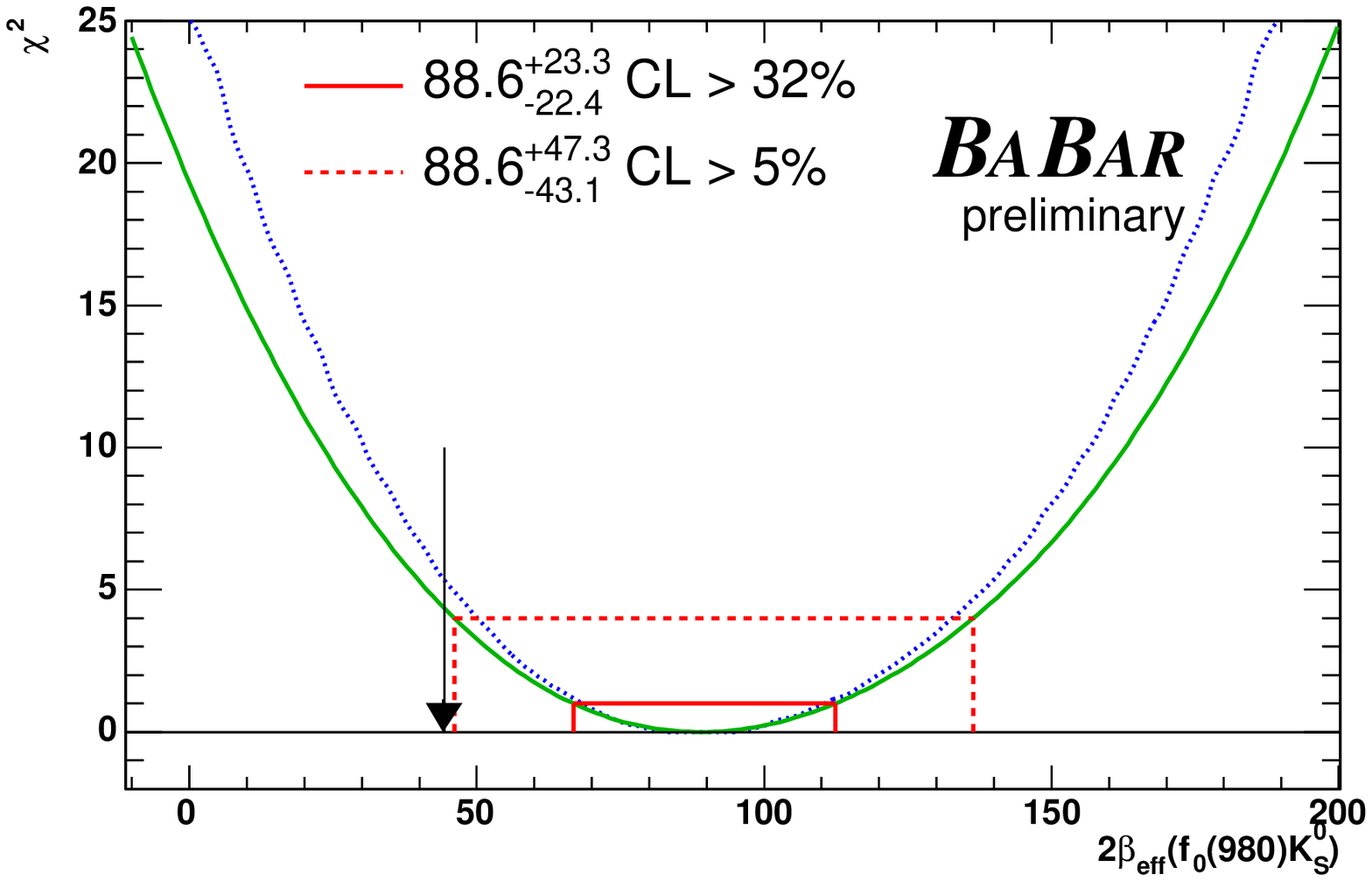}
 \epsfxsize9cm\epsffile{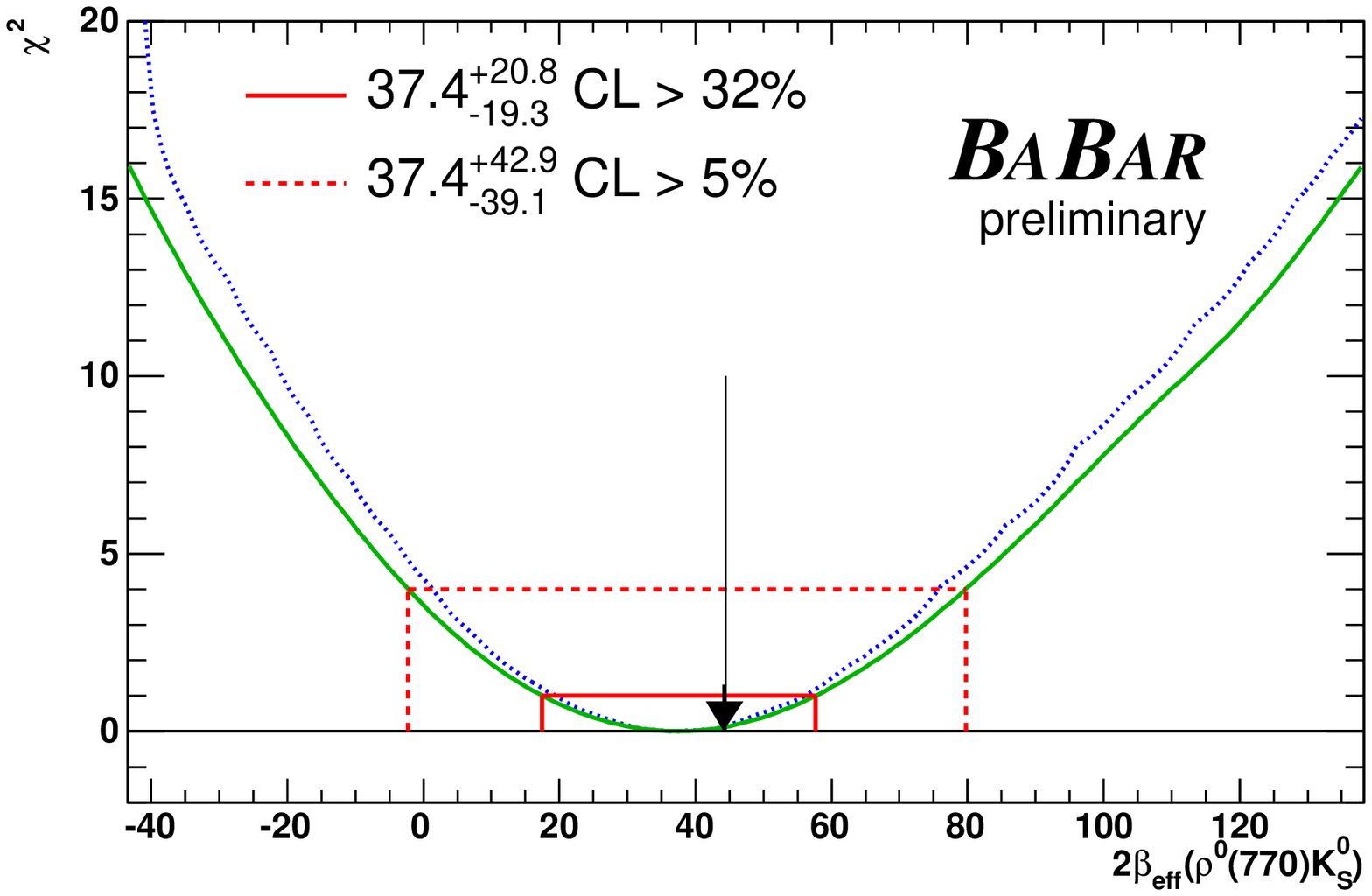}}
  \centerline{  \epsfxsize9cm\epsffile{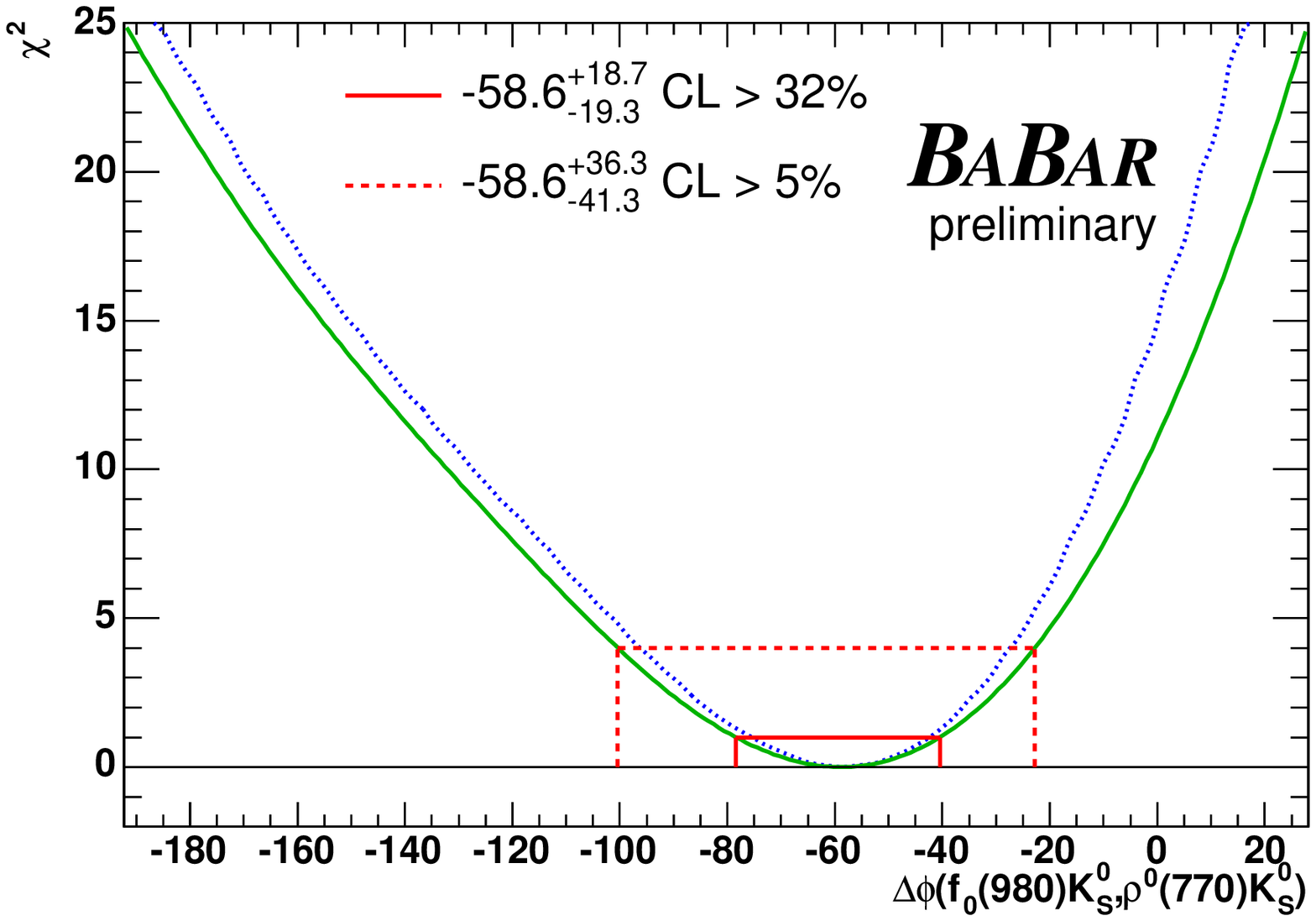}
 \epsfxsize9cm\epsffile{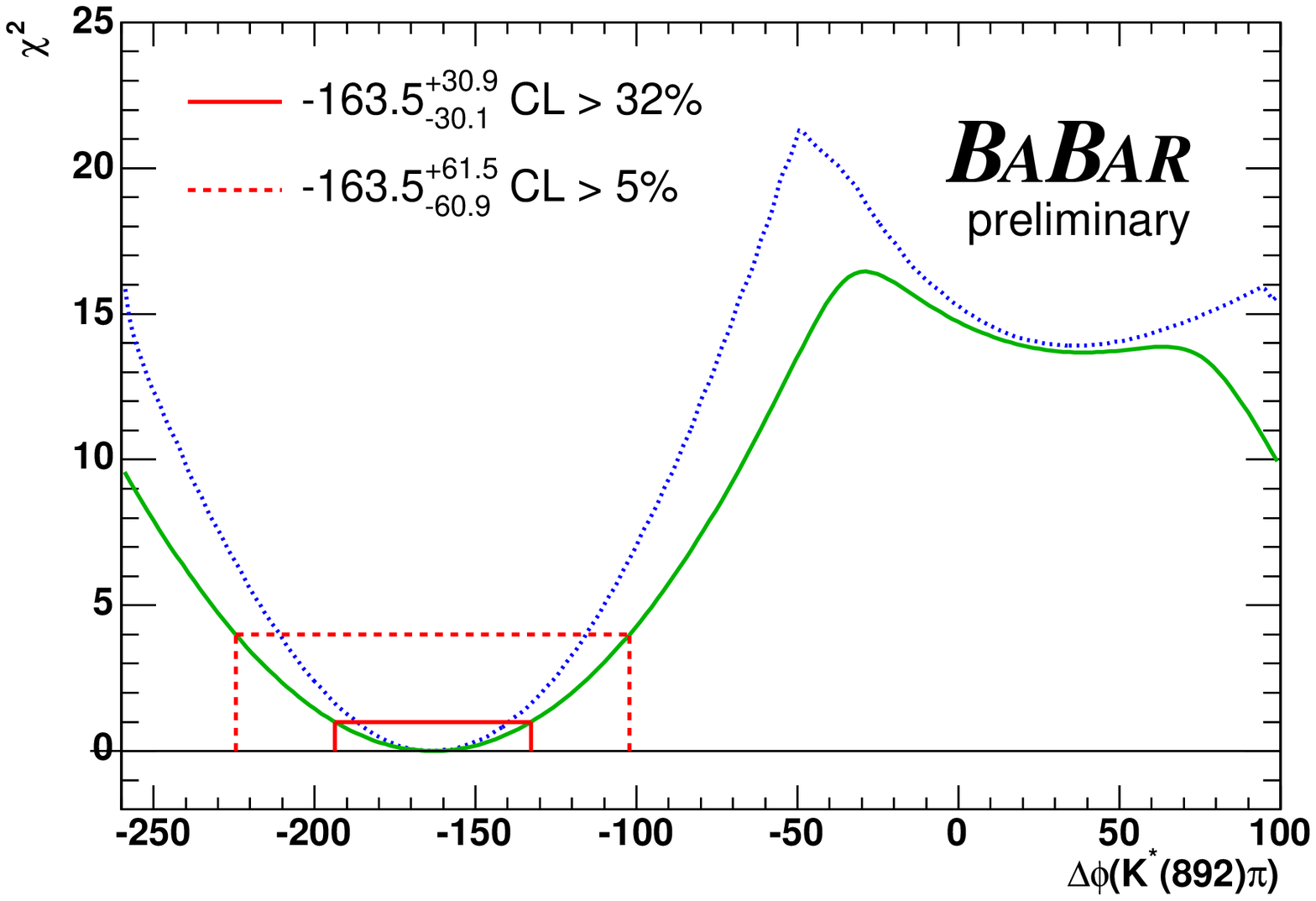}}
  \caption{\em \label{fig:scans} Results of likelihood scans in terms of
        $\chi^2 = -2\ln({\cal L})$ for the Q2B parameters (left to right, top to bottom):
	$S(f_0(980)K^0_S)$, $S(\rho^0(770)K^0_S)$,
        $2\beta_{\rm eff}(f_0(980)K^0_S)$, $2\beta_{\rm eff}(\rho^0(770)K^0_S)$,
        $\Delta\phi(f^0(980)K^0_S,\rho(770)K^0_S)$ and $\Delta\phi(\KstarpI\pim, \KstarmI\pip)$.   
        The solid (dotted) curves corresponds to the total (statistical) error.
        Indicated by solid (dashed) rectangles are the parameter values corresponding to 
        $1\sigma$ ($2\sigma$). The arrows mark the measured values in 
        $b \to c\bar{c}s$ transitions~\cite{hfag}.}
\end{figure}

The Q2B parameters and fit fractions are given together with their statistical
and systematic errors in Table~\ref{tab:resultsQ2B}. It is indicated in the table whether 
the statistical error has been computed by a likelihood scan or using the linear approximation.
Systematic uncertainties are discussed in Section~\ref{sec:Systematics}. Results of 
likelihood scans are shown in Figure~\ref{fig:scans} in terms of $\chi^2 = -2\ln({\cal L})$.

The measured values of $2\beta_{\rm eff}(\fI\KS)$ and $2\beta_{\rm eff}(\rhoI\KS)$ are 
both consistent with the SM predictions. For the former, 
the measured value is higher by $2.1$ standard deviations compared
to that for $b \to c\bar{c}s$. This is unlike the tendency of other results in $b \to q\bar{q}s$ transitions. In addition to this, $2\beta_{\rm eff}(\fI\KS) = 0$ is excluded at $4.3\,\sigma$ significance.
The phase difference $\Delta\phi(\KstarpI\pim,\,\KstarmI\pip)$ is measured here for the first time.
A mirror solution at $\sim 35^\circ$ is disfavored at $3.7 \sigma$ significance. The interval 
$-102.0^\circ < \Delta\phi(\KstarpI\pim, \KstarmI\pip) < 135.7^\circ$
is excluded at $95\%$ confidence level (CL).

Note that the values of the Q2B parameters $S$ and $C$ that are obtained from our fit variables must take
values within the physical boundary $S^2 + C^2 < 1$, in contrast to the values obtained from Q2B analysis.
The value of $S(\fI\KS)$ that is obtained is close to the physical boundary and consequently has a highly
non-Gaussian uncertainty, as shown in terms of $\chi^2$ in Figure~\ref{fig:scans}. The positive uncertainties
on $S(\fI\KS)$, including systematics, extracted from this distribution are $0.09$ at the $32\%$ CL and $0.31$ 
at the $5\%$ CL.  The systematic error on $S$ is derived from the corresponding errors on $2\beta_{\rm eff}$ and 
$C$. 

To validate the presence of the $\fX\KS$, $\fII\KS$ and $\chi_{c0}(1P)\KS$ resonant modes in the signal, we performed fits with three reduced signal models, in which we removed these modes one by one. The differences between these fits and the nominal one in terms of $\chi^2$ for the models missing $\fX\KS$, $\fII\KS$ and $\chi_{c0}(1P)\KS$ are $74$, $37$ and $35$, respectively.

As a validation of our treatment of the time-dependence  we allow
$\tau_{\Bz}$ and $\Delta m_d$ to vary in the fit. We find
$\tau_{\Bz} = ( 1.579 \pm 0.061) \ps$ and 
$\Delta m_d = ( 0.497 \pm 0.035 ) \ps^{-1}$
while the remaining free parameters are consistent with the nominal fit.
In addition, we performed a fit floating the $S$ parameters for
$\Bz\rar J/\psi \KS$ and $\Bz\rar \psi(2S) \KS$ events.  We 
measure $S=\sin(2\beta)=0.690\pm0.077$ and  $0.73\pm0.27$ for 
$J/\psi\KS$ and $\psi(2S) \KS$ respectively.  These numbers are
in agreement with the current world average for $\sin(2\beta)$. 
To validate the SCF modeling, we leave the average SCF fractions per tagging
category free to vary in the fit and find results that are consistent
with the MC estimation. 


\section{SYSTEMATIC STUDIES}
\label{sec:Systematics}

\begin{table*}[t]
\begin{center}
\setlength{\tabcolsep}{0.0pc}
\begin{tabular*}{\textwidth}{@{\extracolsep{\fill}}lccc}
\hline
 \rule[-6pt]{0pt}{18pt}  & $C(\fI\KS)$ & $2\beta_{\rm eff}(\fI\KS)$  & $f(\fI\KS)[\%]$ \\
\hline\\[-0.35cm]
 DP model  	         &  0.04  &  7.8  &  0.6  \\
 Lineshape parameters    &  0.06  &  3.9  &  1.0  \\
 \B background           &  0.03  &  2.7  &  0.4  \\
 Fit bias                &  0.01  &  1.1  &  1.0  \\
 Other                   &  0.03  &  1.4  &  0.1  \\
\hline\\[-0.3cm]
     Sum w/o DP model    & 0.07   &  5.1  &  1.5  \\
\hline\\[-0.3cm]
   Total sum             &  0.08  &  9.3  &  1.6  \\
\hline
\end{tabular*}
\vspace{1.5\baselineskip}

\vspace{-0.35cm}
\setlength{\tabcolsep}{0.0pc}
\begin{tabular*}{\textwidth}{@{\extracolsep{\fill}}lccc}
\hline
 \rule[-6pt]{0pt}{18pt} & $C(\rhoI\KS)$ & $2\beta_{\rm eff}(\rhoI\KS)$ & $f(\rhoI\KS)[\%]$ \\
\hline\\[-0.35cm]
 DP model           	  &  0.06  &  5.9   &  1.1 \\
 Lineshape parameters     &  0.04  &  3.6   &  0.3 \\
  \B background          &  0.06  &  3.8   &  0.1 \\
  Fit bias               &  0.02  &  0.4   &  1.0 \\
  Other                  &  0.02  &  1.0   &  0.1 \\
\hline\\[-0.3cm]
     Sum w/o DP model    &  0.08  &  5.3    &  1.1 \\
\hline\\[-0.3cm]
   Total sum            &  0.10  &  7.9    &  1.5 \\
\hline
\end{tabular*}
\vspace{1.5\baselineskip}

\vspace{-0.35cm}
\setlength{\tabcolsep}{0.0pc}
\begin{tabular*}{\textwidth}{@{\extracolsep{\fill}}lcccc}
\hline
 \rule[-6pt]{0pt}{18pt} & $A_{\CP}(K^{*+}\pim)$ & $f(K^{*+}\pim)[\%]$ & $\Delta\phi(\Kstarp\pim, \Kstarm\pip)$ & $\Delta\phi(f_0\KS,\rho^0\KS)$  \\
\hline\\[-0.35cm]
 DP model              &  0.03 &  0.6  &  15.0   &  6.0 \\
 Lineshape parameters  &  0.01 &  0.2  &  4.3    &  4.2 \\
 \B background         &  0.03 &  0.3  &  4.5    &  4.3	\\
 Fit bias    	         &  0.01 &  1.2  &  9.7    &  0.3	\\
 Other                 &  0.00 &  0.1  &  2.6    &  1.7	\\
\hline\\[-0.3cm]
     Sum w/o DP model  &  0.03 &  1.3  &  11.8   &  6.3 \\
\hline\\[-0.3cm]
   Total sum           &  0.05 &  1.4  &  19.1   &  8.7 \\
\hline
\end{tabular*}
\vspace{1.5\baselineskip}

\vspace{-0.35cm}

\vspace{-0.2cm}
\caption{ \label{tab:systematics}
        \em Summary of systematic uncertainties. Errors on $2\beta_{\rm eff}$ and $\Delta \phi$ are given in degrees and relative fractions in $\%$. $\Kstarpm$ refer to $\KstarpmI$.}
\end{center}
\end{table*}

The contributions to the systematic error on the signal parameters are 
summarized in Table~\ref{tab:systematics}. 

To estimate the contribution to $\BztoKspipi$ decay via other
resonances, we have first
fitted the data including these other decays in the fit model.
We considered possible resonances,
including  $\mbox{$\omega(782)$}$, $\rhoII$, $\mbox{$\rhoz(1700)$}$,
$f_0(1710)$, $f_2(1810)$, $\KstarpmIV$, $K^{*\pm}_2(1430)$,
$\chi_{c2}(1P)$
and a low mass $S$ wave $\sigma$. A RBW lineshape has been used to parameterize these additional
resonances, with masses and widths from~\cite{pdg2006}.
As a second step we have simulated high statistic samples of events, using a model based on the previous fits, including the additional resonances. Finally, we fitted these simulated samples using the nominal signal model.
The systematic effect (contained in the ``DP model'' field in 
Table~\ref{tab:systematics}) is estimated by observing the difference
between the generated values and the fitted values corresponding to the generated samples.
This systematic effect is quoted separately. 

We vary the mass, width and other parameters (if any) of all components in the fit
within their errors, as quoted in Table~\ref{tab:model}, and assign the observed
differences in the measured amplitudes as systematic uncertainties
(``lineshape parameters'' in Table~\ref{tab:systematics}). 

To validate the fitting tool, we perform fits on large MC samples with
the measured proportions of signal, continuum and $B$ background events.
No significant biases are observed in these fits. The statistical
uncertainties on the fit parameters are taken as systematic uncertainties
(``Fit bias'' in Table~\ref{tab:systematics}).

Another major source of systematic uncertainty is the $B$ background model. 
The expected event yields from the background modes are varied according 
to the uncertainties in the measured or estimated branching fractions.  
Since $B$ background modes may exhibit  \CP violation, the corresponding 
parameters are varied within their uncertainties, or if unknown, within the physical range.
As is done for the signal PDFs, we vary the $\dt$ resolution parameters and
the flavor-tagging parameters within their uncertainties and assign
the differences observed in the  data fit with respect to the 
nominal fit as systematic errors.  
The systematic uncertainties from these sources 
are listed as  ``\B Background '' in Table~\ref{tab:systematics}.  

Other systematic effects are much less important for the measurements
of the amplitudes and  are combined in 
the ``Other'' field in Table~\ref{tab:systematics}. Details are given
below.

The parameters of continuum PDFs are determined by the fit. No additional systematic 
uncertainties are assigned to them. An exception to this is the DP
PDF: to estimate the systematic
uncertainty from the $\mes$ sideband extrapolation, we select large 
samples of off-resonance data by loosening the requirements on $\de$ and 
the NN output. We compare the distributions of $\mprime$ and $\thetaprime$ 
between the $\mes$ sideband and the signal region. No significant 
differences are found. We assign as systematic error the effect seen when
weighting the continuum DP PDF by the ratio of both data 
sets. This effect is mostly statistical in origin. 

The uncertainties associated with $\dmd$ and $\tau$ are
estimated by varying these parameters within the uncertainties
on the world average~\cite{pdg2006}.

The signal PDFs for the $\dt$ resolution and tagging fractions 
are determined from fits to a control sample
of fully reconstructed \B decays to exclusive final states with
charm, and the uncertainties are obtained by varying the parameters
within the statistical uncertainties.

The average fraction of misreconstructed signal events predicted by the MC
simulation has been verified with fully reconstructed $\B\to D\rho$
events~\cite{rhopipaper}. No significant differences between data and
the simulation were found. We vary $\fscfave$ for all tagging categories
relatively by $25\%$ to estimate the systematic uncertainty.
Tagging efficiencies, dilutions and biases for signal events
are varied within their experimental uncertainties.

\section{SUMMARY}
\label{sec:Summary}

We have presented preliminary results from a time-dependent Dalitz plot analysis of $\BztoKspipi$ decays
obtained from a data sample of $383$ million $\FourS \to B\Bbar$ decays.
We measure $15$ pairs of relative phases and magnitudes for the different resonances,
taking advantage of the interference between them in the Dalitz plot. 
From the measured decay amplitudes, we derive the Q2B parameters of the resonant decay modes.
In particular, the mixing-induced \CP asymmetry $S$ is extracted from the measured amplitudes.
The measured values of $2\beta_{\rm eff}$ in $\Bz$ decays to $\fI\KS$ and $\rhoI\KS$ are 
$(89^{+22}_{-20}\pm 5 \pm 8)^\circ$ and $(37^{+19}_{-17}\pm 5 \pm 6)^\circ$, respectively. 
These results are both consistent with the SM predictions, but in the case of $\Bz \to \fI\KS$ 
the measured value is higher by $2.1$ standard deviations compared
to that for $b \to c\bar{c}s$. This is unlike the tendency of other results in $b \to q\bar{q}s$ transitions. 
Also, $2\beta_{\rm eff}(\fI\KS) = 0$ is excluded at $4.3\,\sigma$ signficance. 

In decays to $\KstarI \pi$ we find $A_{\CP} = -0.18 \pm 0.10\pm 0.03 \pm 0.03$.
The phase difference $\Delta\phi$ between the amplitudes of $\Bz \to \KstarpI \pim$ and 
$\Bzb \to \KstarmI \pip$ is measured for the first time.
We find $\Delta\phi = (-164\pm 24 \pm 12 \pm 15)^\circ$, and mirror solutions are disfavored 
at $\sim 3.7 \sigma$ significance. The interval 
$-102^\circ < \Delta\phi < 136^\circ$ is excluded at $95\%$ confidence level. 
Our results may be used to extract the CKM angle $\gamma$ following the methods proposed in Refs.~\cite{Deshpande:2002be,Ciuchini:2006kv,Gronau:2006qn,Lipkin:1991st}.

\section{Acknowledgments}
\label{sec:acknowledgments}
We are grateful for the 
extraordinary contributions of our \pep2\ colleagues in
achieving the excellent luminosity and machine conditions
that have made this work possible.
The success of this project also relies critically on the 
expertise and dedication of the computing organizations that 
support \babar.
The collaborating institutions wish to thank 
SLAC for its support and the kind hospitality extended to them. 
This work is supported by the
US Department of Energy
and National Science Foundation, the
Natural Sciences and Engineering Research Council (Canada),
the Commissariat \`a l'Energie Atomique and
Institut National de Physique Nucl\'eaire et de Physique des Particules
(France), the
Bundesministerium f\"ur Bildung und Forschung and
Deutsche Forschungsgemeinschaft
(Germany), the
Istituto Nazionale di Fisica Nucleare (Italy),
the Foundation for Fundamental Research on Matter (The Netherlands),
the Research Council of Norway, the
Ministry of Science and Technology of the Russian Federation, 
Ministerio de Educaci\'on y Ciencia (Spain), and the
Science and Technology Facilities Council (United Kingdom).
Individuals have received support from 
the Marie-Curie IEF program (European Union) and
the A. P. Sloan Foundation.

\clearpage
\bibliography{leppho2007}
\bibliographystyle{leppho2007}

\end{document}